\documentclass[aoas,preprint]{imsart}

\usepackage{graphicx}
\usepackage{natbib}
\usepackage{here, url}
\usepackage{multirow}
\usepackage{amsmath, amssymb}
\usepackage{bm}
\usepackage{subfigure}
\usepackage{color}
\usepackage{lineno}

\begin{document}
\begin{frontmatter}


\title
{Bayesian factor models for probabilistic cause of death assessment with verbal autopsies}

\begin{aug}
\author{\fnms{Tsuyoshi} \snm{Kunihama}\thanksref{m1}\ead[label=e1]{t.kunihama@kwansei.ac.jp}},
\author{\fnms{Zehang Richard} \snm{Li}\thanksref{m2}\ead[label=e2]{lizehang@gmail.com}}
\author{\fnms{Samuel J.} \snm{Clark}\thanksref{m3}\ead[label=e3]{work@samclark.net}}
\and
\author{\fnms{Tyler H.} \snm{McCormick}\thanksref{m4}
\ead[label=e4]{tylermc@uw.edu}
\ead[label=u1,url]{http://www.stat.washington.edu.com}}

\runauthor{Kunihama et al.}

\affiliation{Kwansei Gakuin University\thanksmark{m1}, Yale University\thanksmark{m2}, the Ohio State University\thanksmark{m3}, and University of Washington\thanksmark{m4}}

\address{Kwansei Gakuin University\\
Nishinomiya, Japan\\
\printead{e1}}

\address{University of Washington\\
Seattle, Washington USA\\
\printead{e2}\\
\phantom{E-mail:\ }\printead*{e4}}

\address{{T}he Ohio State University\\
Columbus, Ohio USA\\
\printead{e3}}
\end{aug}





\begin{abstract}

The distribution of deaths by cause provides crucial information for public health planning, response, and evaluation. About 60\% of deaths globally are not registered or given a cause, limiting our ability to understand disease epidemiology. Verbal autopsy (VA) surveys are increasingly used in such settings to collect information on the signs, symptoms, and medical history of people who have recently died. This article develops a novel Bayesian method for estimation of population distributions of deaths by cause using verbal autopsy data. The proposed approach is based on a multivariate probit model where associations among items in questionnaires are flexibly induced by latent factors. 
Using the Population Health Metrics Research Consortium labeled data that include both VA and medically certified causes of death, we assess performance of the proposed method.  Further, we estimate important questionnaire items that are highly associated with causes of death. This framework provides insights that will simplify future data collection.  \\

\end{abstract}

\begin{keyword}
\kwd{Bayesian latent model}
\kwd{cause of death}
\kwd{conditional dependence}
\kwd{multivariate data}
\kwd{verbal autopsies}
\kwd{survey data}
\end{keyword}

\end{frontmatter}

\section{Introduction}

The distribution of causes of death is an essential part of understanding population dynamics, as well as implementing and evaluating effective public health interventions \citep[e.g.][]{Ruzicka90, Mathers05,soleman2006verbal,bloomberg2015understanding}. Monitoring cause of death requires understanding both acute, rapid onset epidemiological crises (e.g. the outbreak of an infectious diseases) and observing changes over the course of decades (e.g. the rise in obesity related diabetes). The ``gold-standard" for assigning cause of death relies on physical autopsies with pathological reports. In low-resource settings, however, most deaths happen outside of hospitals and are often not recorded by a civil registration and vital statistics systems \citep{mikkelsen2015global}.  In such settings, understanding the mortality burden of a specific cause, and trends in cause-specific mortality over time, is extremely challenging \citep[e.g.][]{phillips2015well,de2017integrating}.  In-person autopsies take valuable physician time away from patients to perform autopsies and are extremely difficult or impossible for deaths that happen outside of the hospital, meaning that any insights are generalizable only to the small fraction of the population that regularly interacts with the healthcare system.

Scaling up to a system that can record all deaths presents massive financial and logistical challenges, meaning that local and national governments must rely on lower cost alternatives.  One common approach is to sue surveys with relatives or caretakers of the decedent.  There is, naturally, much less information in a survey-based system than in a physical autopsy and there are numerous data quality challenges.  Nonetheless, given the lack of credible alternatives, survey-based data are  and will continue to be vital for understanding cause of death distributions \citep{Horton07, AbouZahr07,jha2014reliable}.  Survey-based data for cause of death assessment are known as verbal autopsies (VAs) and consist of interviews with a family member or other individual familiar with the death.  The respondent answers a questionnaire about the signs, symptoms, demographic characteristics and health history of the deceased individual. Deaths are typically identified using community informants or using a partial surveillance system. Interviews are conducted by specially trained enumerators, some of whom have medical expertise. VA surveys are widely conducted \citep{Lopez98, Yang05, Maher10, Sankoh12}, and the World Health Organization (WHO) releases a standardized VA questionnaire to facilitate comparison across areas \citep{whoStandardTools2012,whoStandardTools2016,nichols20182016}. 

Given VA surveys, there are several available methods to estimate a cause of death based on the reported symptoms.  In some settings, trained clinicians review VAs and assess a cause of death \citep{Lozano11}.  This approach can be effective in some circumstances, but is time-consuming and requires that trained clinicians (many of whom would otherwise be seeing patients) be available. An alternative approach is to use an algorithmic or statistical method to assign causes of death.  Several such methods have been proposed and evaluated in the statistics and public health literatures  \citep[see for example][]{James11,Byass12,Serina15,Miasnikof15,McCormick16}.   

For the most part, these methods rely on a critical assumption: independence across symptoms conditional on a given cause.  This assumption disregards critical information about constellations or clusters of symptoms that are typical of a given cause and thus particularly informative when assigning causes.  The only method currently available that uses information about dependence between symptoms is work by King and Lu~\citep{King08,King10}. The King and Lu method regresses the probability of random subsets of symptoms on the conditional probability of the selected symptoms given a cause.  This process is an attempt to represent the space of all possible symptom combinations.  However since there are typically one to two hundred symptoms, exploring all possible combinations is an extremely daunting task, which is often impossible in reasonable time.

Our work presents a novel approach to incorporating dependence between symptoms in assigning cause of death from VA surveys.  In our approach, we capture dependence between symptoms using a small number of latent factors.  This approach avoids the need to evaluate all possible symptom combinations as in the King and Lu framework.  We build a multivariate probit model for symptoms conditional on a cause. Binary-scale outcomes can be interpreted as a manifestation of underlying continuous variables.  A factor model on these conditional variables provides a sparse covariance structure between symptoms. Our method also accommodates missing data that commonly arise in VA surveys, because for example, family members may not remember all details about sign/symptoms of the deceased person.  The proposed approach can incorporate both individual-specific and design-based missing values by summing them out from the probit model with a missing-at-random assumption. We fit the model using an efficient Markov chain Monte Carlo (MCMC) algorithm we develop for posterior computation. Further, we utilize our framework to better understand the importance of each measure in the questionnaire.  To do this, we quantify the association between each symptom with each cause, using a model-based version of Cram{\' e}r's $V$.  Our measures can be used to simplify and shorten future VA surveys, decreasing both the burden on respondents and the cost.

The remainder of the paper is organized as follows. This section describes labeled VA data from the Population Health Metrics Research Consortium (PHMRC) that will be the primary data source we use in our analysis. Section 2 proposes a novel approach for estimation of population distributions of causes of death using a small number of latent factors. Section 3 develops an efficient MCMC algorithm. Section 4 assesses the performance of the proposed approach in various scenarios and measures strength of dependence of questionnaire items in the PHMRC dataset. Section 5 concludes the article.

\subsection{PHMRC VA survey} \label{sec:data}

The Population Health Metrics Research Consortium (PHMRC) collected VA data at six study sites in four countries: Andhra Pradesh, India; Bohol, Philippines; Dar es Salaam, Tanzania; Mexico City, Mexico; Pemba Island, Tanzania; and Uttar Pradesh, India. In each study site, VAs were collected for adults, children and neonates in hospital and clinical environments. Causes were assigned based on diagnostic criteria including laboratory, pathology and medical imaging findings. \citet{Murray11b} provide the detailed criteria for each cause, which were developed by a committee of physicians involved in the study.  The cause list was constructed based on WHO global burden of disease estimates of the leading causes of death in the developing world.  VA interviews were conducted with a relative of the deceased by interviewers who were blinded to the cause of death assigned in the hospital, and the family member provided the consent for the VA study. The VA questionnaire items cover symptoms of illnesses, demographic characteristics, diagnoses of chronic illnesses by health service providers, possible risk factors such as tobacco-use and other potentially contributing characteristics. In typical settings where VA surveys are implemented, it is not possible to obtain a large fraction of deaths with physician codes. The PHMRC data are, therefore, intended to be used as training data for statistical and algorithmic methods used in settings where only VA surveys are available. \citet{Murray11b} provide detailed information of the design and implementation of the PHMRC study, and \citet{data} released a version of the dataset to the public after removing potentially identifying information from the original VA interviews. The lists of causes of death and predictors in our analysis are in the supplementary materials.

Figure \ref{fig:hist-cod} shows the barplot of 34 pre-defined causes of death for adults in the PHMRC data, and those for each study site are in the supplementary materials. We observe that there is a large difference in the number of deaths between causes, for example 630 in stroke and 40 in esophageal cancer, and distributions of the causes vary considerably among the sites.  The VA questionnaire consists of binary, count and categorical items.  Existing statistical and algorithmic tools for assigning cause of death from VA surveys dichotomize categorical and continuous variables. We use the procedure described in \citet{Murray11b} and \citet{McCormick16} to convert all indicators into binary variables, leading to a dataset with 7,841 individuals and 175 symptoms/indicators. Dichotomizing continuous and categorical variables no doubt loses some information, but it also facilitates greater comparability with current state of the art methods for VA classification. Building models for high dimensional, mixed-scale variables remains a challenging open area of research. An additional feature of the PHMRC, and all VA data, is the presence of abundant missing values.  These occur for several reasons, including difficulty recalling specific circumstances of a person's death. Figure \ref{fig:boxplot} shows the histogram of the missing rate of the binary predictors. We observe many predictors contain missing values and in some cases the missingness is extreme, with more than half of questions missing answers.

\begin{figure}[ht] 
\begin{center}
\includegraphics[scale=0.85]{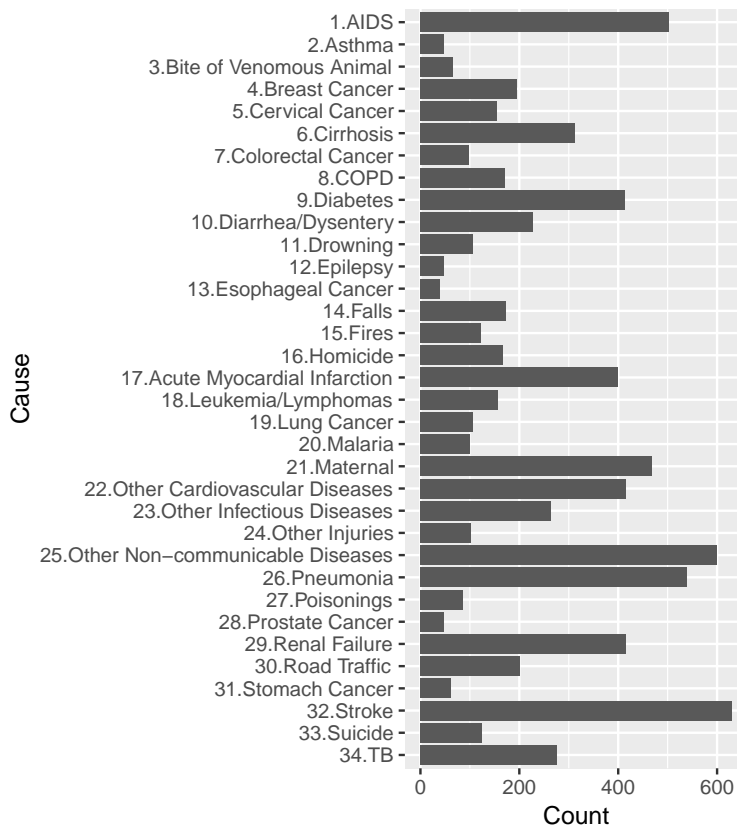}
\caption{Barplot of 34 causes of death in the PHMRC data. We follow \citet{data} in the numbering of the causes.}
\label{fig:hist-cod}
\end{center}
\end{figure}

Since the PHMRC data contain medically-certified causes, we can explore the magnitude of the dependence between symptoms for a given medically-certified cause. We compute Cram{\' e}r's $V$ \citep{cramer46} that measures strength of associations between two variables, taking a value from 0 (no association) to 1 (complete association). Figure 3 shows the result for all pairs of predictors for deaths due to AIDS related causes. We conducted chi-squared test using the R function {\tt cramersV} in the {\tt lsr} package \citep{lsr, R}, and the hypothesis of independence was rejected with 5\% significance level for 1669 pairs of the predictors out of 
$15225$ 
pairs. Using Fisher's exact test, the hypothesis was rejected with 5\% significance level for 1843 pairs. Unlike most of the previously available methods, our proposed method will utilize these correlations to improve cause assignment accuracy.

\begin{figure}[htpb] 
\begin{center}
\includegraphics[scale=1]{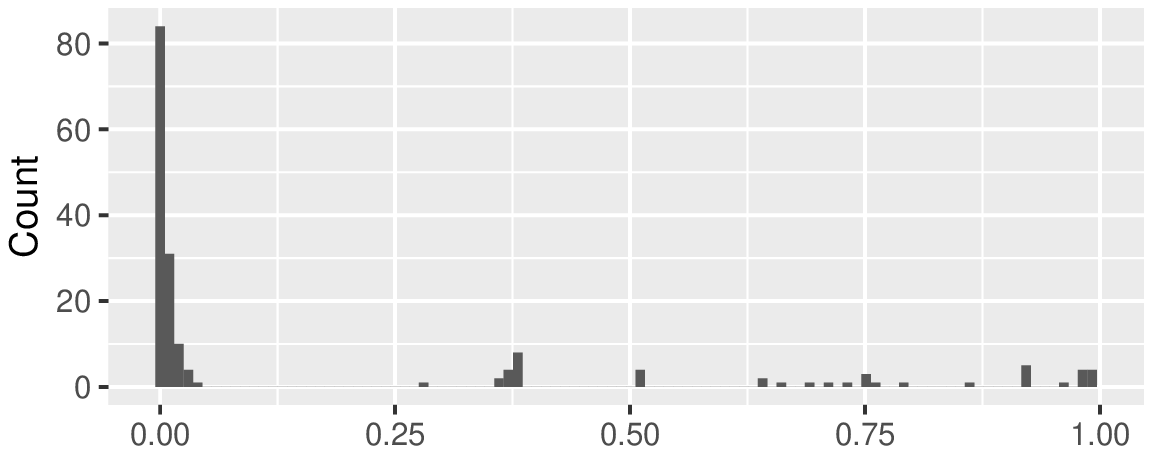}
\caption{Histogram of the missing rate of the predictors. The horizontal and vertical axes show the missing rate and counts of the predictors.}
\label{fig:boxplot}
\includegraphics[scale=0.9]{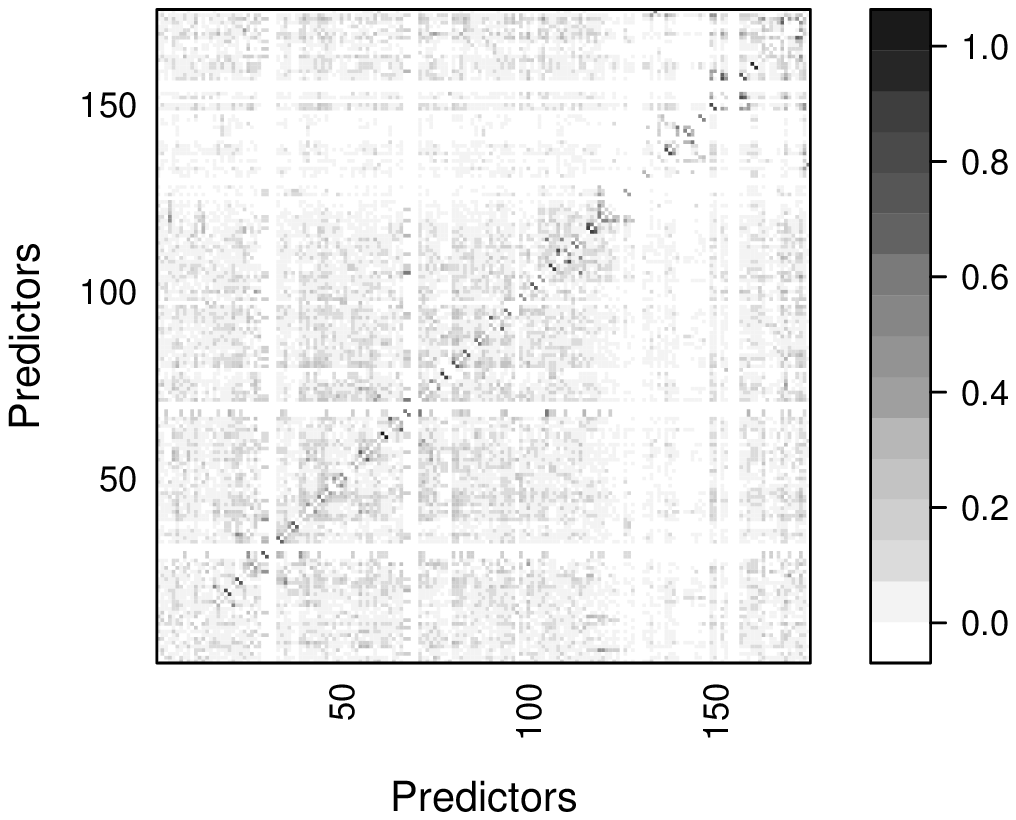}
\caption{Cram{\' e}r's V between predictors for the deaths caused by AIDS. The horizontal and vertical axes present the index of each predictor. Details of the predictors are in the supplementary materials. The scale on the right hand side shows the value of Cram{\' e}r's V.}
\label{fig:corr}
\end{center}
\end{figure}



\section{Bayesian factor model for VA data}
In this section we present our model formulation.  First we present the modeling framework, and then we discuss how to assess the importance of symptoms.  

\subsection{Bayesian approach} \label{sec:model}

We propose a novel Bayesian framework for assessing cause of death using VA surveys. Let $y_i \in \{1,\ldots, C\}$ be the cause of death of the $i$th person with $i=1,\ldots,n$, and $x_i=(x_{i1}, \ldots, x_{ip})'$ be the responses to questions $j=1,\ldots,p$ with $x_{ij} \in \{0, 1\}$. One approach would be to directly build a conditional probability of a person having died from a certain condition given a set of observable covariates $\pi(y_i\,|\, x_i)$ using standard parametric models (e.g. multinomial probit/logit regressions). Modeling these conditional probabilities directly is unappealing, however, since we have a high fraction of missing data (see Figure~\ref{fig:boxplot}), we would also need to impute a substantial fraction of symptoms. Performing these imputations would require choosing an imputation model, which would bring with it particular assumptions about the structure of the data.  Overall, it is not straightforward to build a flexible model for high dimensional binary data with complex interactions, and there is a possibility that the imputation model may fail to capture an actual structure of the data.

Given the challenges with imputing missing values, we opt instead for a Bayesian framework where we can integrate over missing symptoms.  To do this, we first express the conditional distribution, $\pi(y_i \,|\,x_i)$, using Bayes' rule:
\begin{align*}
\pi(y_i \,|\,x_i) = \frac{ \pi(x_i\,|\, y_i) \pi(y_i) }{ \sum_{c=1}^C \pi(x_i\,|\, y_i = c) \pi( y_i = c)   }.
\end{align*}
In this framework we can integrate out missing values from $\pi(x_i\,|\,y_i)$. Let $x_i^{obs}$ and $x_i^{mis}$ denote the observed and missing items for the $i$th person with $x_i = (x_i^{obs}, x_i^{mis})$. We assume missing data are {\it missing at random}, i.e., the probability of the missing-data mechanism depends on observed data but not on the missing values \citep{Rubin76, Seaman13, Fabrizia15}. Under this assumption, one can conduct inference on parameters in a model using only observed information. 
We utilize the distribution of the cause given the observed items,
\begin{align*}
\pi(y_i \,|\,x^{obs}_i) \propto \pi(x^{obs}_i\,|\, y_i) \pi(y_i),\mbox{ where } \pi(x^{obs}_i\,|\, y_i) = \int  \pi(x_i\,|\, y_i) dx_i^{mis}.
\end{align*}
If the integral can be calculated analytically, we can evaluate the conditional probabilities of the cause on the observed information without imputing missing data. Without assuming {\it missing at random}, an additional model is needed to describe the missing-data mechanism.  
In the case of VA data, however, we do not have the requisite information.  Missing data could arise because respondents were asked a question but did not recall the answer.  This is expected since the interview is about a traumatic event, requires recalling specialized details, and often does not happen until months after the death.  Missing data could also arise because interviewers, perhaps with an eye towards a likely cause, strategically asked certain questions or asked questions in a specific order.  An active area of research in the VA community involves gathering ethnographic and numeric data to describe the VA interview process, with the hopes of making the process more standard across contexts and understanding missing data mechanisms. Utilizing information about the interview and respondent recall processes, to construct a missing data model still remains an open area of research. 
Therefore, in the absence of an alternative missing data model, we can enhance the plausibility of the assumption by incorporating as many variables as possible in a model \citep{little2002statistical, gelman2013bayesian}.

Under the Bayes' rule representation, there are two pieces of the model that we need to specify, (i) the unconditional distribution of individual causes, $\pi(y_i)$, and (ii) the conditional distribution of observed symptoms given an individual has a particular cause, $\pi(x_i\,|\, y_i)$.  Beginning with the prior distribution for causes, we assume a Dirichlet distribution,
\begin{align*}
\left\{ \pi(y_i=1), \ldots, \pi(y_i=C) \right\} \sim \text{Dirichlet}(a_1,\ldots,a_C)
\end{align*}
where $a_1,\ldots,a_C$ are concentration parameters. Since cause patterns can differ substantially across geographic areas and times, we assume we have little prior information about the distribution of causes.  We therefore assume $a_1=\cdots=a_C=1$, leading to a uniform prior with $\pi(y_i=c) \propto 1$ for $c=1,\ldots,C$.  

The second piece, $\pi(x_i\,|\, y_i)$, requires modeling a set of high-dimensional binary predictors given each cause.  As described previously, nearly all existing methods for assigning cause of death from verbal autopsies make the assumption of conditional independence across symptoms \citep{Byass12, Miasnikof15, McCormick16},
\begin{align*}
\pi(x_i \,|\, y_i) = \prod_{j=1}^p \pi(x_{ij} \,|\, y_i).
\end{align*}
This assumption facilitates computation but disregards substantial and potentially informative relationships between symptoms, as Figure~\ref{fig:corr} shows.  

To flexibly capture dependence, we develop a conditional distribution based on the multivariate probit model.  In our framework, each binary outcome is a manifestation of an underlying continuous variable.  Let $z_i=(z_{i1},\ldots,z_{ip})' \in \mathbb{R}^p$ be the latent variable for the $i$th person. We express the multivariate binary variable $x_i$ by transforming the continuous variable $z_i$. We assume a multivariate normal distribution conditional on a cause, $z_i \,|\, y_i \sim N(\mu_{y_i}, \Sigma_{y_i})$ with mean $\mu_{y_i} = (\mu_{y_i 1},\ldots,\mu_{y_i p})'$ and covariance $\Sigma_{y_i}$. 

Even for moderately large $p$, estimating $p(p+1)/2$ parameters in the covariance matrix will be challenging, particularly since we expect that each dataset will contain only a few deaths by each cause.  Further, since our goal is predicting cause of death for a new sample of deaths, we prefer a sparse model to avoid overfitting.  Rather than estimating all elements in the covariance matrix, we introduce a $K$-dimensional factor $\eta_i = (\eta_{i1},\ldots,\eta_{iK})'$ with $K \ll p$, and propose the following sparse factor model,
\begin{align}
x_{ij} &= 1( z_{ij} > 0 ), \ \ j=1,\ldots,p, \nonumber \\
z_i &= \mu_{y_i} + \Lambda_{y_i} \eta_i + \varepsilon_i, \ \ \eta_i \sim N(0,I_K), \ \ \varepsilon_i \sim N( 0, I_p ),
\label{eq:1}
\end{align}
where $1(\cdot)$ is an indicator function and $\Lambda_y=\{ \lambda_{yjk} \}$ is a $p\times K$ loading matrix with $y=1,\ldots,C$, $j=1,\ldots,p$ and $k=1,\ldots,K$.  Dependence is induced in $z_i$ by integrating out the factor $\eta_i$ in (\ref{eq:1}), leading to the normal distribution with cause-dependent mean and covariance,
\begin{align*}
z_i \,|\, y_i \sim N( \mu_{y_i},  \Lambda_{y_i} \Lambda'_{y_i} + I_p ),
\end{align*}
where the number of parameters in the covariance reduces from $p(p+1)/2$ to $Kp$. In practice, we will need to choose the number of factors, $K$, which we propose doing via cross validation. For the prior distribution of the mean and factor loadings, we use Cauchy distributions, a standard shrinkage prior with high density around zero and heavy tails.  The Cauchy prior reduces effects of redundant elements, but will also capture meaningful signals. Based on the normal-gamma distribution, we express the Cauchy distribution as
\begin{align*}
\mu_{yj} &\sim N(0, \tau^{-1}_{j}), \ \ \tau_j \sim Ga(0.5, 0.5), \\
\lambda_{yjk} &\sim N(0, \phi^{-1}_{j}), \ \ \phi_j \sim Ga(0.5, 0.5),
\end{align*}
where $Ga(a,b)$ denotes the gamma distribution with mean $a/b$. The latent variables $\tau_j$ and $\phi_j$ are shared among the causes and factors for reduction of the number of  parameters in the model. 

In a factor model, constraints on the factor loadings are necessary to identify the latent factors \citep{Arminger1998, West04, Zhou14}. In our context, though, identification is not necessary since the factors are a way to reduce the dimension of the covariance matrix, rather than to be interpreted in and of themselves.  We opt not to use constraints for identification since interpreting the factors is not a goal and constraints can lead to order dependence and computational inefficiencies \citep{Bhattacharya11, Montagna12}.

\subsection{Measuring strength of association} \label{sec:measure}

VA surveys collect information via questionnaires with many items regarding demographic background, health history and disease symptoms. It can be time-consuming and costly in terms of both enumerator time and the toll on a person close to the decedent to ask redundant questions that will not be useful in predicting likely cause of death.  
We now propose a method for assessing the importance of each symptom based on the strength of association with causes of death. 

For two discrete random variables $y\in\{1,\ldots,m_y\}$ and $x\in\{1,\ldots,m_x\}$, \citet{Dunson09} develop a Bayesian measure of association relying on the parameters which characterize the multivariate distribution,
\begin{align}
\delta^2 = \frac{1}{ \min \{m_y, m_x \} - 1 }  \sum_{c=1}^{m_y} \sum_{d=1}^{m_x} \frac{ \{ P( y = c, x = d ) - P( y = c ) P( x = d ) \}^2 }{ P( y = c ) P( x = d )  },
\label{eq:approx}
\end{align}
where $\delta$ ranges from 0 to 1 with $\delta \approx 0$ if $y$ and $x$ are independent. If the joint and marginal probabilities are replaced by the empirical distributions, $\delta$ will correspond to Cram{\' e}r's $V$. Therefore, $\delta$ can be considered a model-based version of Cram{\' e}r's $V$. In the proposed model, it is straightforward to evaluate the probability functions in (\ref{eq:approx}).  For example, the joint probability of the cause and $j$th predictor is expressed as $\pi(y_i,x_{ij}) = \pi(x_{ij}\,|\,y_i) \pi(y_i)$ where $\pi(x_{ij}\,|\,y_i)$ is the proposed factor model and $\pi(y_i)$ is the Dirichlet distribution.  In the following section, we describe how the measure can be incorporated into our posterior sampling algorithm.

\section{Posterior computation} \label{sec:mcmc}

The posterior density for the model presented in Section~\ref{sec:model} is not available in closed form.  We instead approximate the posterior density using samples obtained through Markov-chain Monte Carlo (MCMC). Let $m_i = (m_{i1},\ldots,m_{ip})'$ be a vector of indicators denoting missing values for the $i$th person such that $m_{ij}=1$ if $x_{ij}$ is missing and $m_{ij}=0$ if $x_{ij}$ is observed with $j=1,\ldots,p$. We define notation $[m_i]$ such that, for a vector $b$ and a matrix $B$ with $p$ rows, $b_{[m_i]}$ and $B_{[m_i]}$ denote the subvector and submatrix consisting of components with $m_{ij}=0$ for $j=1,\ldots,p$. Then we propose the following MCMC algorithm. 

\begin{enumerate}

\item Update $\mu_{\cdot j} \equiv (\mu_{1j}, \ldots, \mu_{Cj})'$ from $N(\mu_*, \Sigma_*)$ for $j=1,\ldots,p$ with
\begin{align*}
\mu_* = \Sigma_* a_j, \ \ \Sigma_* = \text{diag}\left\{ (n_1 + \tau_j)^{-1} , \ldots, (n_C + \tau_j)^{-1} \right\},
\end{align*} 
where $n_c = \sum_{i=1}^n 1(y_i = c, m_i = 0)$ and $a_j$ is the $C \times 1$ vector with the $c$th element $\sum_{i=1}^n 1(y_i=c, m_i=0) ( z_{ij} - \lambda_{y_i  j \cdot}' \eta_i )$ where $\lambda_{y_i j \cdot} = (\lambda_{y_i j 1},\ldots, \lambda_{y_i j K} )'$.

\item Update $\lambda_{c j \cdot} \equiv (\lambda_{c j1},\ldots,\lambda_{c jK})'$ from $N(\mu_{\lambda}, \Sigma_{\lambda})$ for $c=1,\ldots, C$ with
\begin{align*}
\mu_{\lambda} = \Sigma_{\lambda} \left\{ \sum_{i: y_i=c} \eta_i ( z_{ij} - \mu_{y_i j} ) \right\}, \ \ \Sigma_{\lambda} = \left( \sum_{i:y_i=c} \eta_i \eta_i' + \phi_j I_K \right)^{-1}.
\end{align*} 

\item Update $\eta_i$ from $N(\tilde{\mu}, \tilde{\Sigma})$ for $i=1,\ldots,n$ with
\begin{align*}
\tilde{\mu} = \tilde{\Sigma} \Lambda'_{y_i [m_i]} (z_i - \mu_{y_i} )_{[m_i]}, \ \ \tilde{\Sigma} = \left( \Lambda'_{y_i [m_i]}  \Lambda_{y_i [m_i]} + I_K \right)^{-1}.
\end{align*} 

\item Update $\tau_j$ for $j=1,\ldots,p$ from
\begin{align*}
Ga \left( \frac{C+ 1}{2}, \frac{\sum_{c=1}^C \mu^2_{c j} + 1}{2}  \right).
\end{align*} 

\item Update $\phi_j$ for $j=1,\ldots,p$ from
\begin{align*}
Ga \left( \frac{CK+ 1}{2}, \frac{ \sum_{c=1}^C \sum_{k=1}^K  \lambda^2_{c jk} + 1}{2}  \right).
\end{align*} 

\item Update $z_{ij}$ with $m_{ij}=0$ for $i=1,\ldots,n$ and $j=1,\ldots,p$ from 
\begin{align*}
\begin{cases}
N_+( \mu_{y_i j} + \lambda_{y_i j \cdot}' \eta_i , 1) & \text{if $x_{ij} =1$}, \\
N_-( \mu_{y_i j} + \lambda_{y_i j \cdot}' \eta_i , 1) & \text{if $x_{ij} = 0$}, \\
\end{cases}
\end{align*}
where $N_+$ and $N_-$ denote the truncated normal distributions with support $[0, \infty)$ and $(-\infty, 0]$ respectively.

\item For a person $i \in S$ where $S$ is the target data, generate $y_i$ with
\begin{align*}
\pi(y_i = c \,|\, x_i^{obs}) = \frac{ \pi(x_i^{obs} \,|\, y_i=c) \pi(y_i=c) }{ \sum_{y=1}^C \pi(x_i^{obs} \,|\, y_i=y) \pi(y_i=y) }, \ \ c=1,\ldots, C,
\end{align*}
where $\pi(x_i^{obs} \,|\, y_i=c) = \int \pi(x_i^{obs} \,|\, \eta, y_i=c) f(\eta) d\eta$ is evaluated using a Monte Carlo approximation with $\eta_r \sim N(0, I_K)$ for $r=1,\ldots,R$,
\begin{align}
\pi(x_i^{obs} \,|\, y_i=c) \approx \frac{1}{R} \sum_{r=1}^R \left\{ \prod_{j:m_{ij}=0} \pi(x_{ij} \,|\, \eta_r, y_i=c) \right\}.
\label{eq:mcmc-r}
\end{align}
Then, compute the population distribution of causes of death by
\begin{align*}
\left( \frac{1}{ \# S } \sum_{i \in S} 1( y_i = 1 ), \ldots, \frac{1}{ \# S } \sum_{i \in S} 1( y_i = C ) \right)
\end{align*}
where $\# S$ is the number of observations in the test data.
\end{enumerate}
For the estimation of strength of dependence in Section \ref{sec:measure}, Step 7 above is replaced by 
\begin{itemize}
\item[7.] Update the distribution of causes $\pi(y_i)$ with Dirichlet($1,\ldots,1$) prior from
\begin{align*}
\text{Dirichlet}\left( \frac{1}{ n } \sum_{i=1}^n 1( y_i = 1 ) + 1, \ldots, \frac{1}{ n } \sum_{i =1}^n 1( y_i = C ) + 1 \right),
\end{align*}
and compute $\delta$ in (\ref{eq:approx}) for each predictor.
\end{itemize}

As mentioned previously, we need to specify the number of latent factors, $K$.  In Section 4, we selected the number of factors by 5-fold cross-validation.  More details are in the supplementary materials. In addition, we set $R=200$ as the number of random samples for the Monte Carlo method in (\ref{eq:mcmc-r}) and generated 5,000 MCMC samples after the initial 500 samples were discarded as a burn-in period, and every 10th sample was saved. Discarding samples in this way is a widely-used approach known as {\it thinning} and is designed to reduce autocorrelations in a Markov chain \citep{Hoff09, gelman2013bayesian}. We observed that the sample paths were stable, and the sample autocorrelations dropped smoothly. Illustrative examples of the sample plot and the autocorrelation are in the supplementary materials. Replication code for the proposed method is publicly available at \url{https://github.com/kunihama/VA-code}.

\section{Results} \label{sec:result}

Using the MCMC algorithm described in the previous section, we fit our model to the PHMRC VA data.  We are particularly interested in the improvement that comes from explicitly accounting for dependence between symptoms. We evaluate whether incorporating dependence between symptoms improves prediction of the distribution of deaths by cause in a target population. As an assessment of the performance, we utilize cause specific mortality fraction (CSMF) accuracy, which is a measure of closeness between two probability vectors in the VA literature \citep{McCormick16},
\begin{align*}
\text{CSMF Accuracy} = 1 - \frac{ d_{L_1}(\pi_0,\pi) }{ 2\{ 1 - \min_{1 \leq c \leq C} \pi_0(y=c) \} },
\end{align*}
where $d_{L_1}$ indicates $L_1$ distance between two distributions and $d_{L_1}(\pi_0, \pi) = \sum_{c=1}^C | \pi_0(y=c) - \pi(y=c) |$.
It is a transformation of $L_1$ distance such that a larger value indicates that two probability vectors are closer and it equals to 1 if they are the same. The true CSMF, $\pi_0$, is approximated by the empirical distribution of causes of death in the test data, and $\pi$ is the estimated distribution of causes of death by a statistical model. In addition, as another measure of the performance, we compute the correlation between the actual and estimated numbers of deaths per cause.

An important consideration in our evaluation is whether the method works when the cause of death distribution (and possibly the relationship between symptoms and causes) varies between the training and testing set. To be clear, we are assuming that the underlying ``true'' population relationship between testing and training remains the same, but that we will have limited training data so our model could be highly sensitive to spurious associations between some symptoms and the rare causes that exist in our training sample. 
This consideration is fundamental in the VA setting since obtaining training data is extremely costly and the fraction of deaths due to each cause can change between testing and training data. For example, in many settings, training data are collected from one geographic area and need to be used for predicting the distribution of deaths by cause at another area. 
In evaluating the method, we consider realistic scenarios where we estimate the distribution of deaths by cause in one area using training data collected from different geographic areas. We study six scenarios in which each of the PHMRC sites is treated as a target site and the rest together as training data. We observe discrepancies in the distributions of deaths by cause between the test and training data in all scenarios. Especially, the case where the test site is Pemba Island shows a relatively large difference in the distributions of causes. The figure is in the supplementary materials.


\begin{figure}[htbp]
  \begin{center}
    \begin{tabular}{c}
      \begin{minipage}{0.5\hsize}
        \begin{center}
          \includegraphics[scale=0.65]{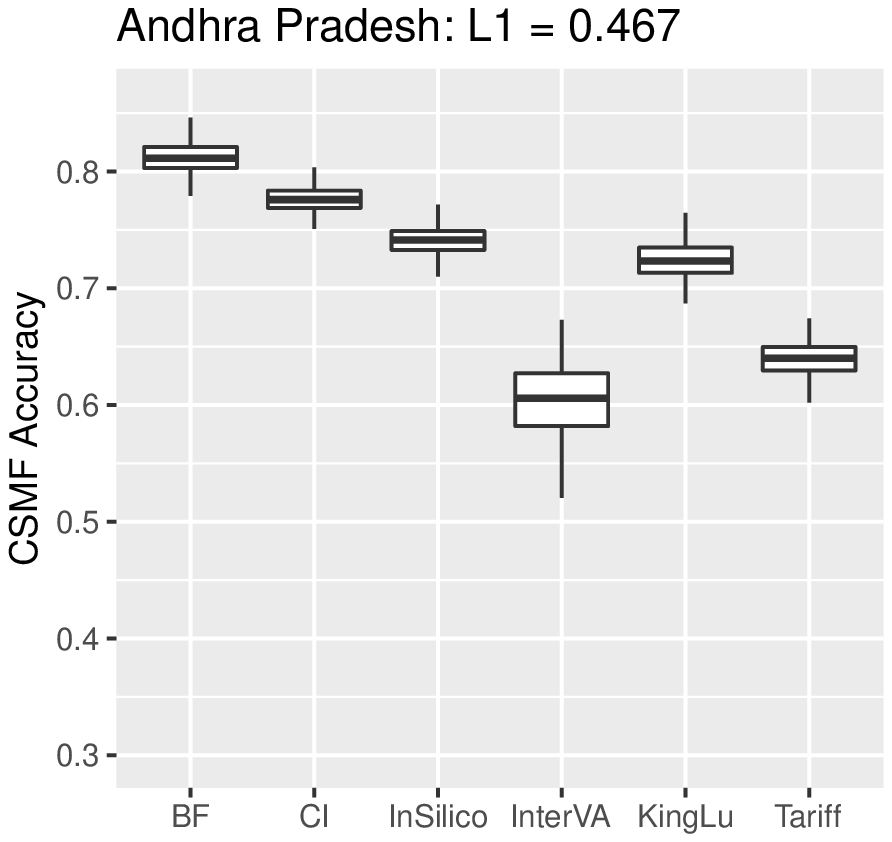}
        \end{center}
      \end{minipage}
      \begin{minipage}{0.5\hsize}
        \begin{center}
          \includegraphics[scale=0.65]{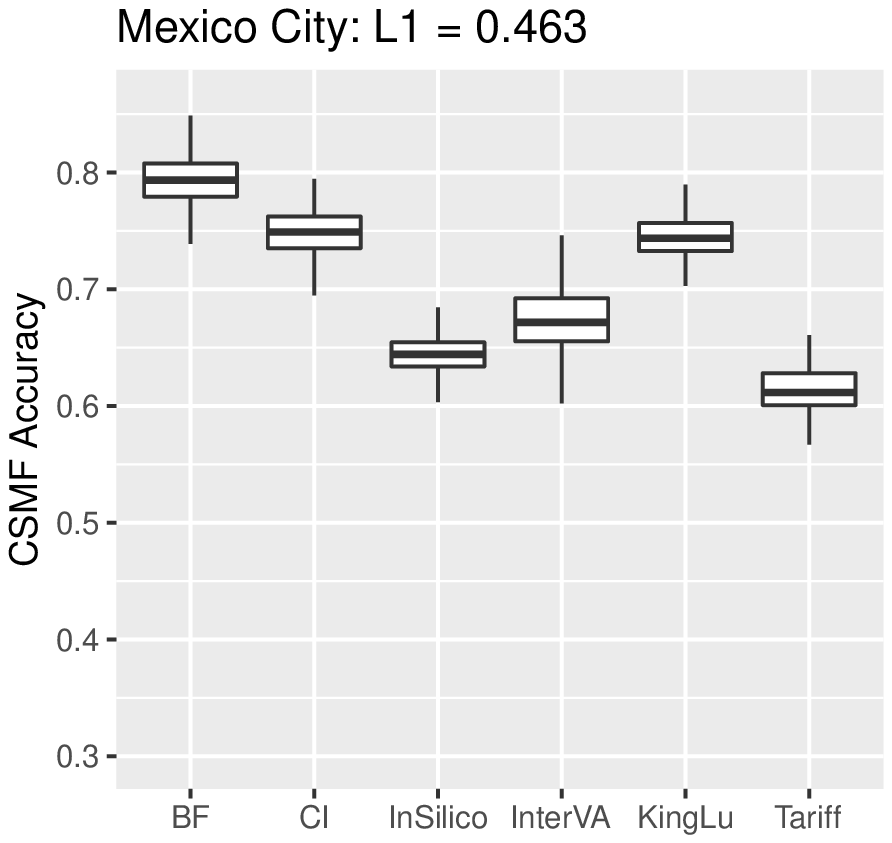}
        \end{center}
      \end{minipage}
    \end{tabular}
    \begin{tabular}{c}
      \begin{minipage}{0.5\hsize}
        \begin{center}
          \includegraphics[scale=0.65]{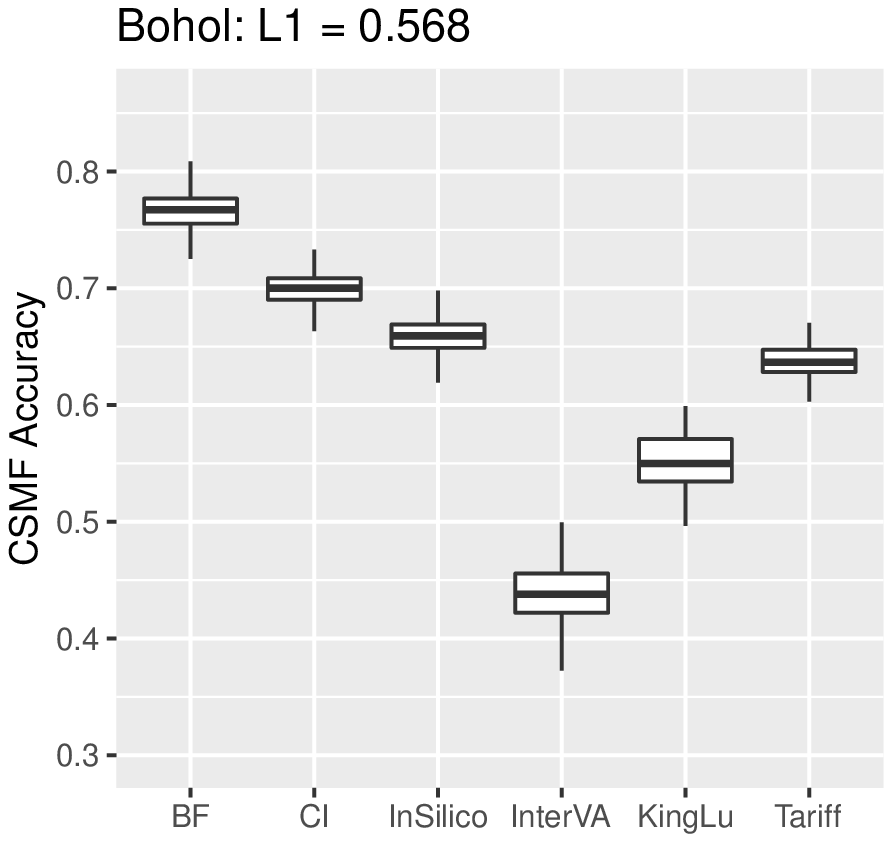}
        \end{center}
      \end{minipage}
      \begin{minipage}{0.5\hsize}
        \begin{center}
          \includegraphics[scale=0.65]{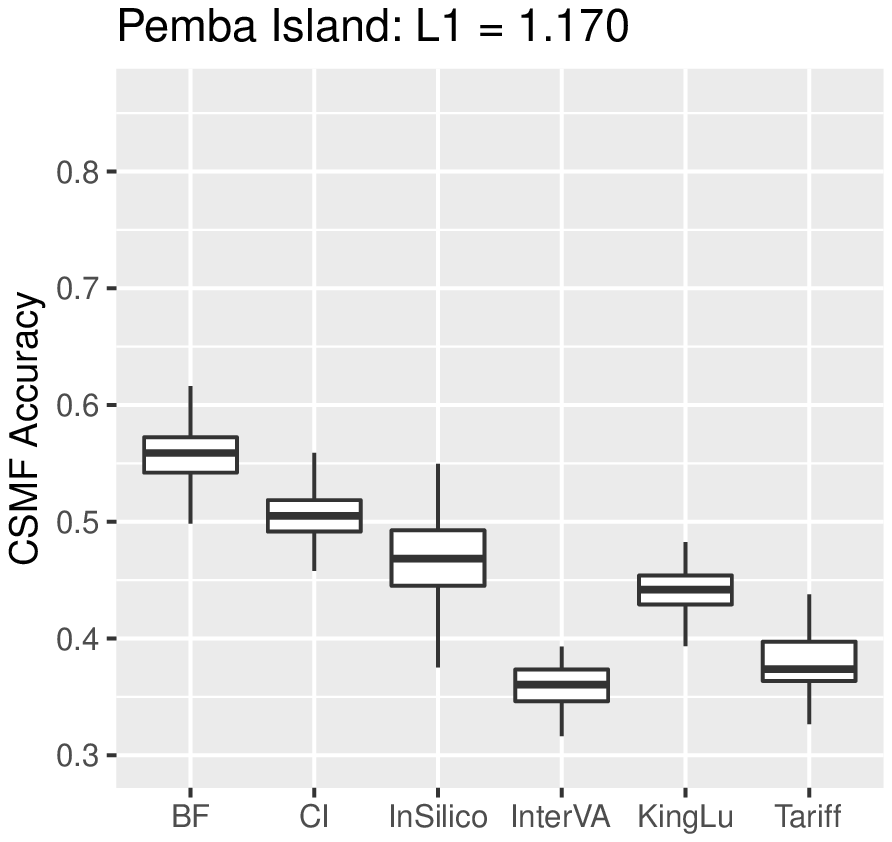}
        \end{center}
      \end{minipage}
    \end{tabular}
    \begin{tabular}{c}
      \begin{minipage}{0.5\hsize}
        \begin{center}
          \includegraphics[scale=0.65]{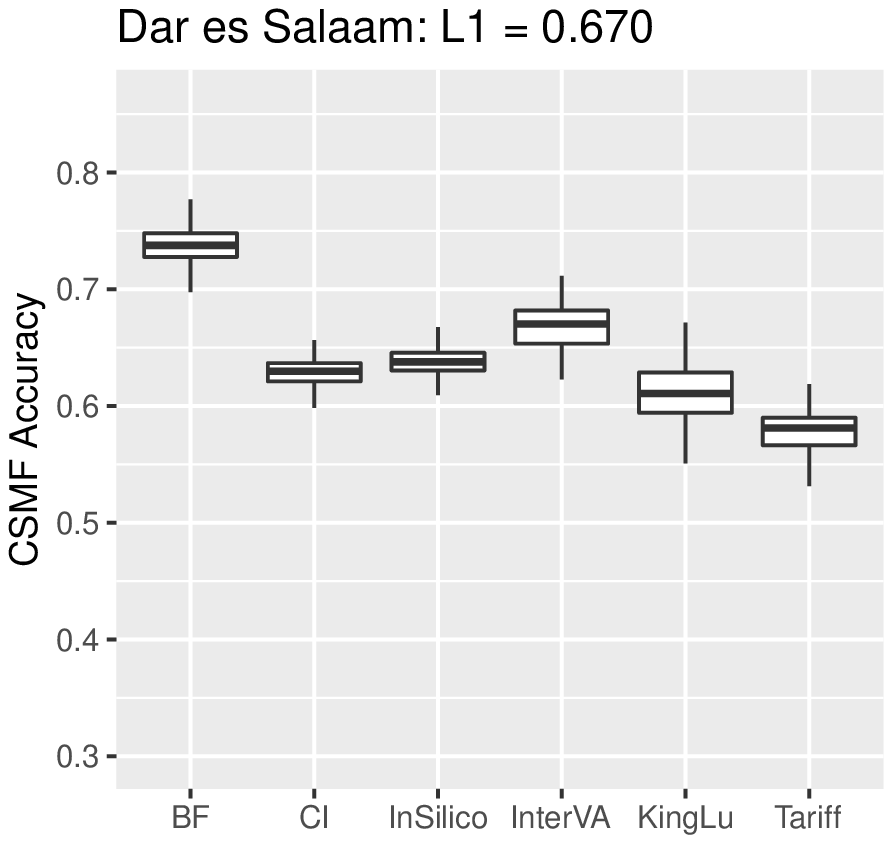}
        \end{center}
      \end{minipage}
      \begin{minipage}{0.5\hsize}
        \begin{center}
          \includegraphics[scale=0.65]{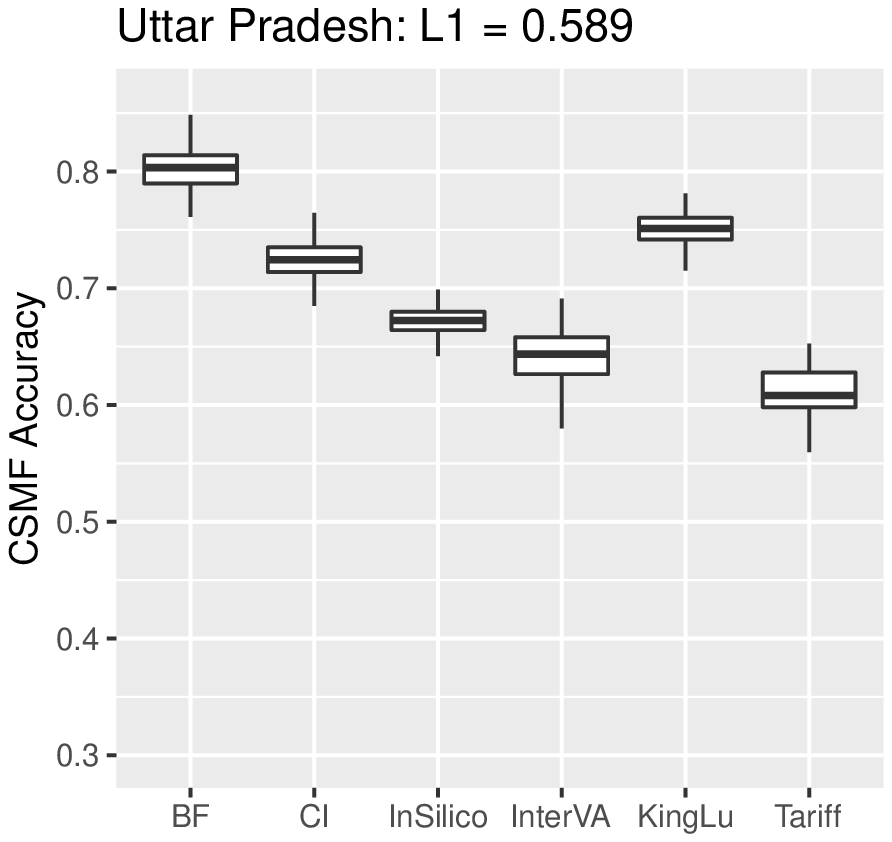}
        \end{center}
      \end{minipage}
    \end{tabular}
\caption{Boxplot of CSMF accuracy by Bayesian factor model (BF), conditionally independent model (CI), InSilicoVA (InSilico), InterVA, King-Lu method (KingLu) and Tariff for scenarios where the target sites are Andhra Pradesh (top left), Bohol (middle left), Dar es Salaam (bottom left), Mexico City (top right), Pemba Island (middle right) and Uttar Pradesh (bottom right). For each scenario, L1 indicates $L_1$ distance between the empirical distributions in the test and training data. A larger $L_1$ distance means a larger difference between the distributions.}
\label{fig:csmf}
\end{center}
\end{figure}

\begin{figure}[htbp]
  \begin{center}
    \begin{tabular}{c}
      \begin{minipage}{0.5\hsize}
        \begin{center}
          \includegraphics[scale=0.65]{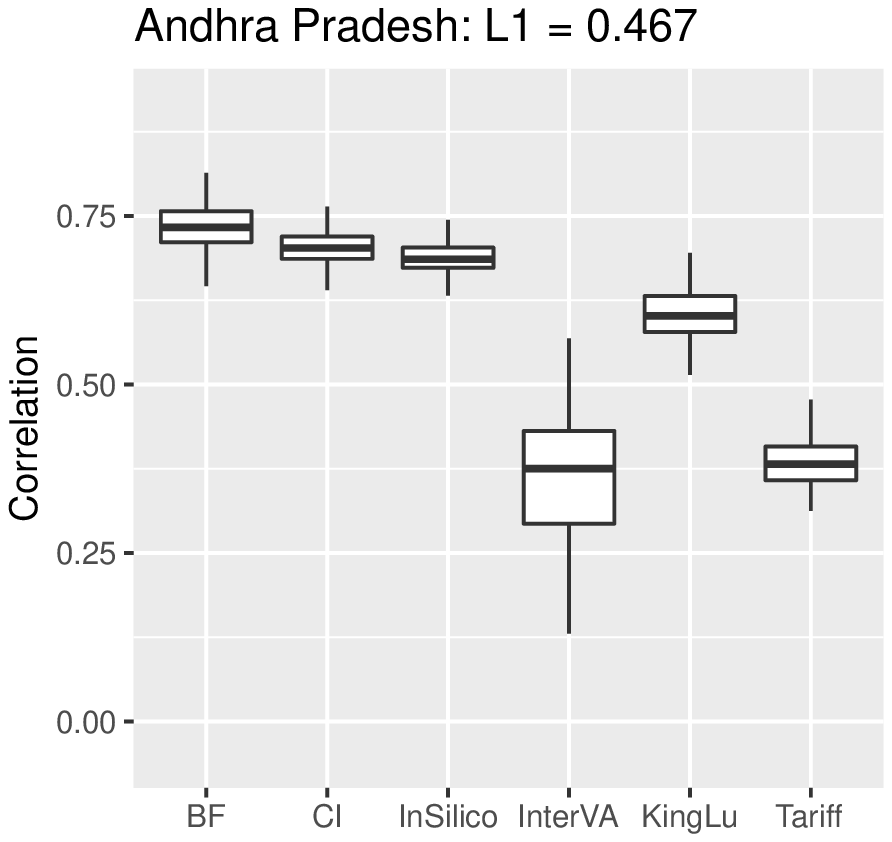}
        \end{center}
      \end{minipage}
      \begin{minipage}{0.5\hsize}
        \begin{center}
          \includegraphics[scale=0.65]{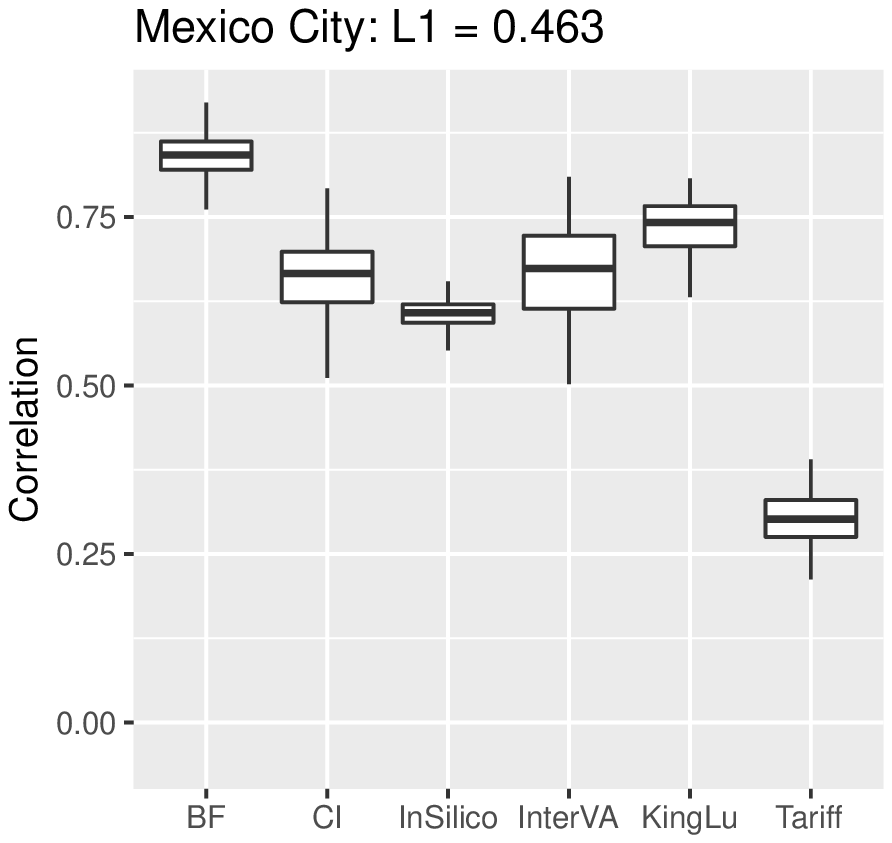}
        \end{center}
      \end{minipage}
    \end{tabular}
    \begin{tabular}{c}
      \begin{minipage}{0.5\hsize}
        \begin{center}
          \includegraphics[scale=0.65]{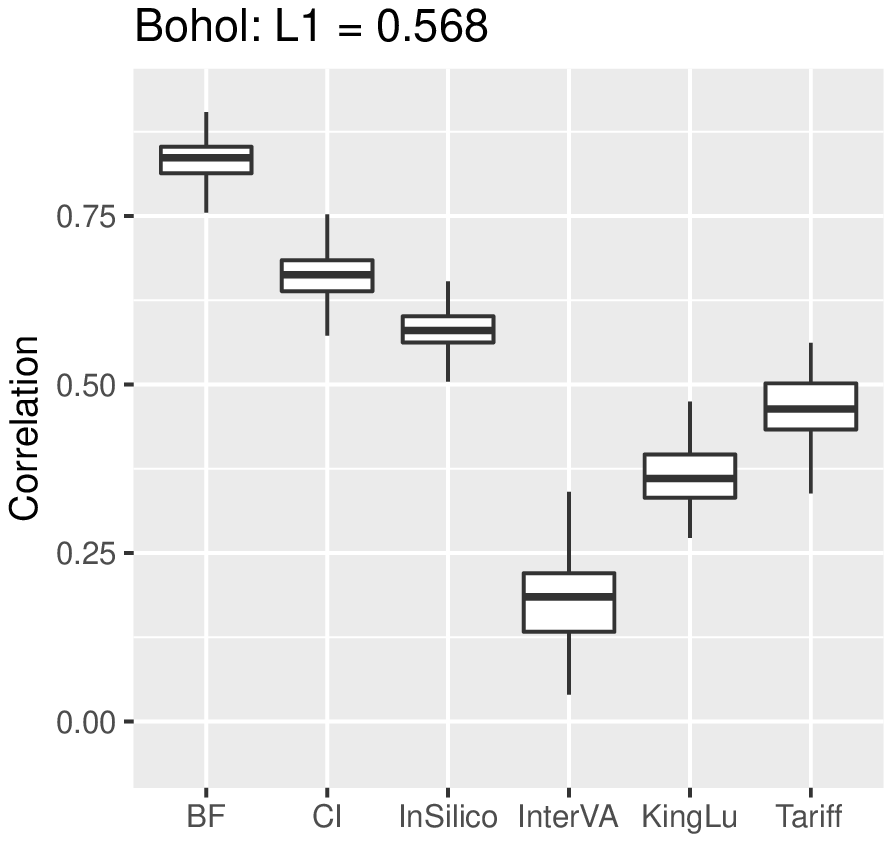}
        \end{center}
      \end{minipage}
      \begin{minipage}{0.5\hsize}
        \begin{center}
          \includegraphics[scale=0.65]{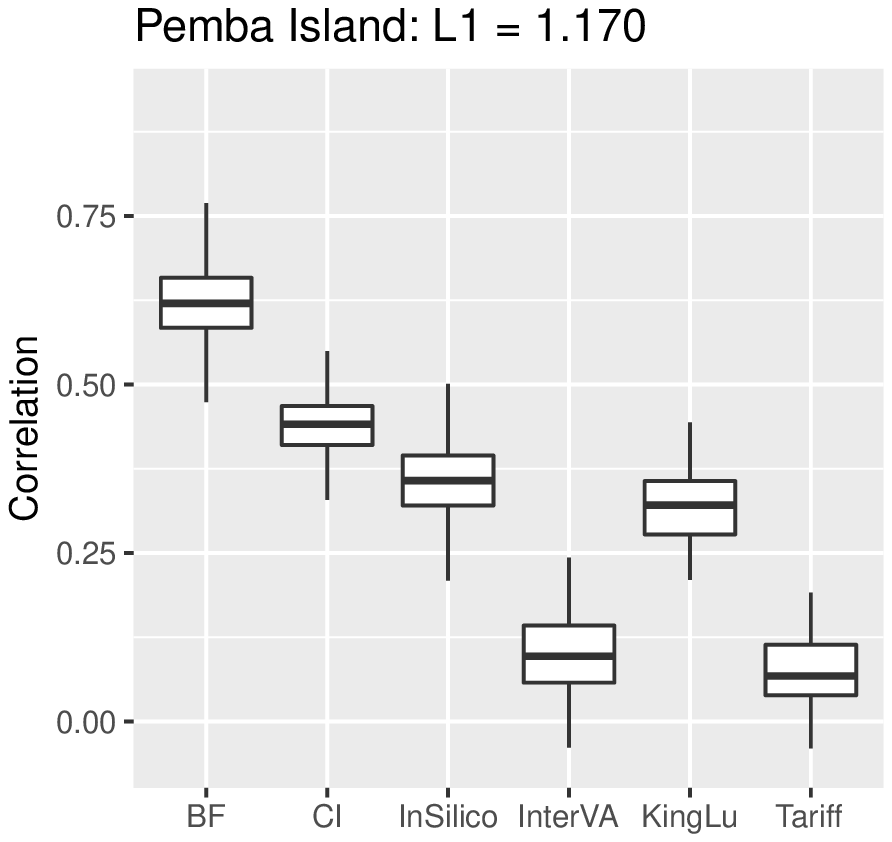}
        \end{center}
      \end{minipage}
    \end{tabular}
    \begin{tabular}{c}
      \begin{minipage}{0.5\hsize}
        \begin{center}
          \includegraphics[scale=0.65]{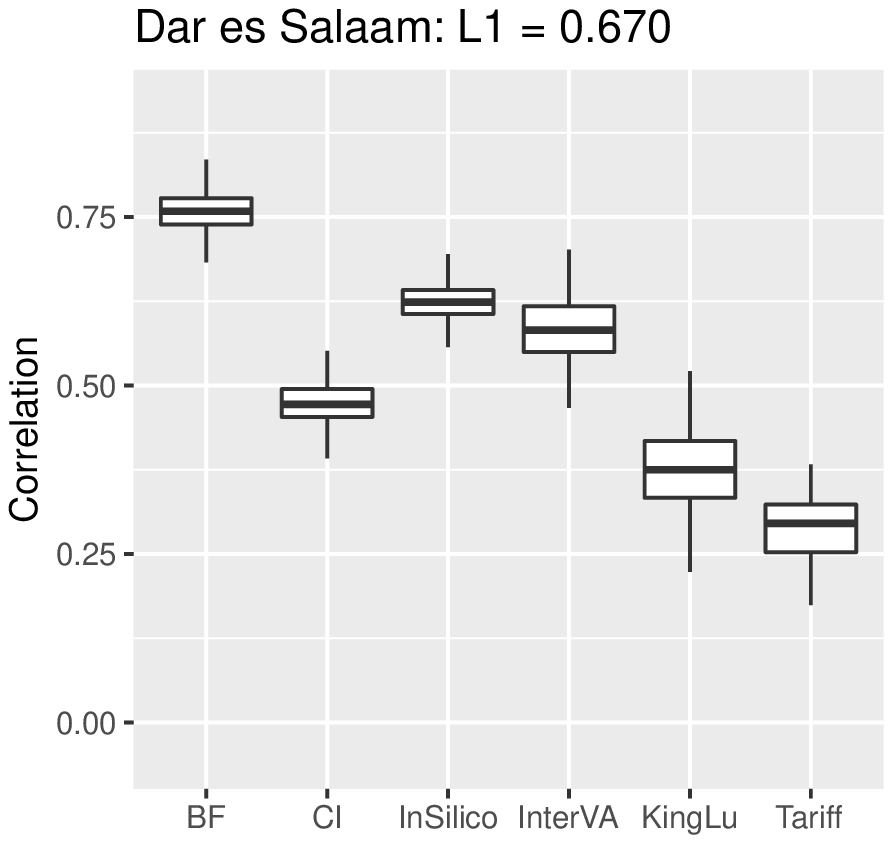}
        \end{center}
      \end{minipage}
      \begin{minipage}{0.5\hsize}
        \begin{center}
          \includegraphics[scale=0.65]{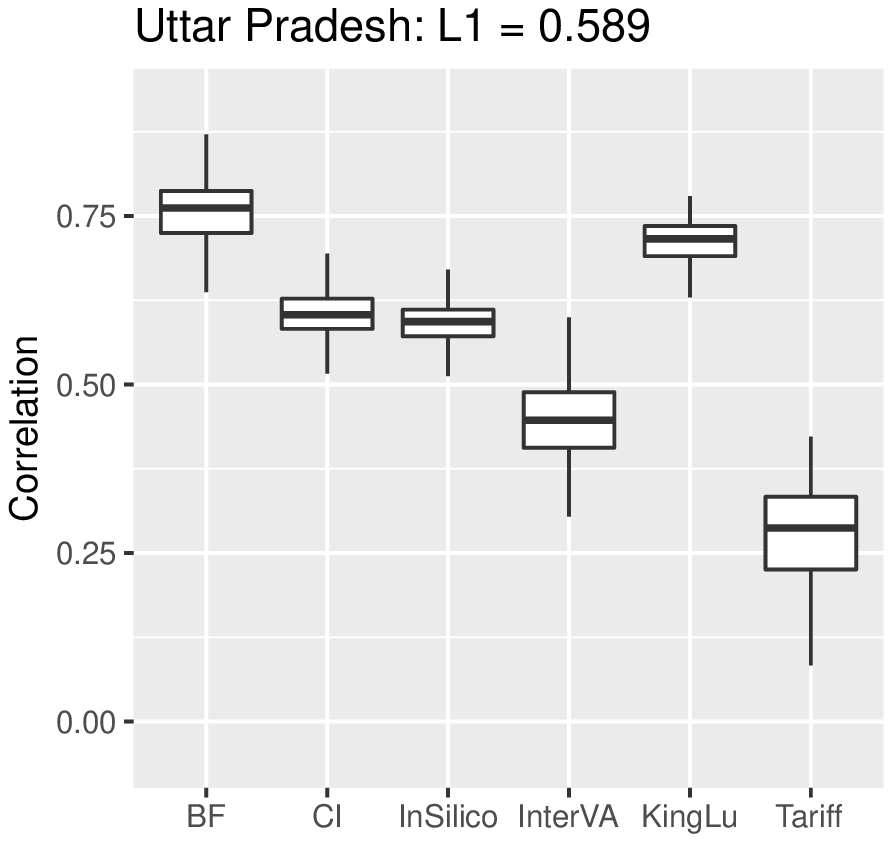}
        \end{center}
      \end{minipage}
    \end{tabular}
\caption{Boxplot of correlation between the actual and predicted numbers of deaths per cause by Bayesian factor model (BF), conditionally independent model (CI), InSilicoVA (InSilico), InterVA, King-Lu method (KingLu) and Tariff for scenarios where the target sites are Andhra Pradesh (top left), Bohol (middle left), Dar es Salaam (bottom left), Mexico City (top right), Pemba Island (middle right) and Uttar Pradesh (bottom right). For each scenario, L1 indicates $L_1$ distance between the empirical distributions in the test and training data. A larger $L_1$ distance means a larger difference between the distributions.}
\label{fig:correlation}
  \end{center}
\end{figure}

\begin{figure}[ht] 
\begin{center}
\hspace{-1.5cm}
\includegraphics[scale=0.88]{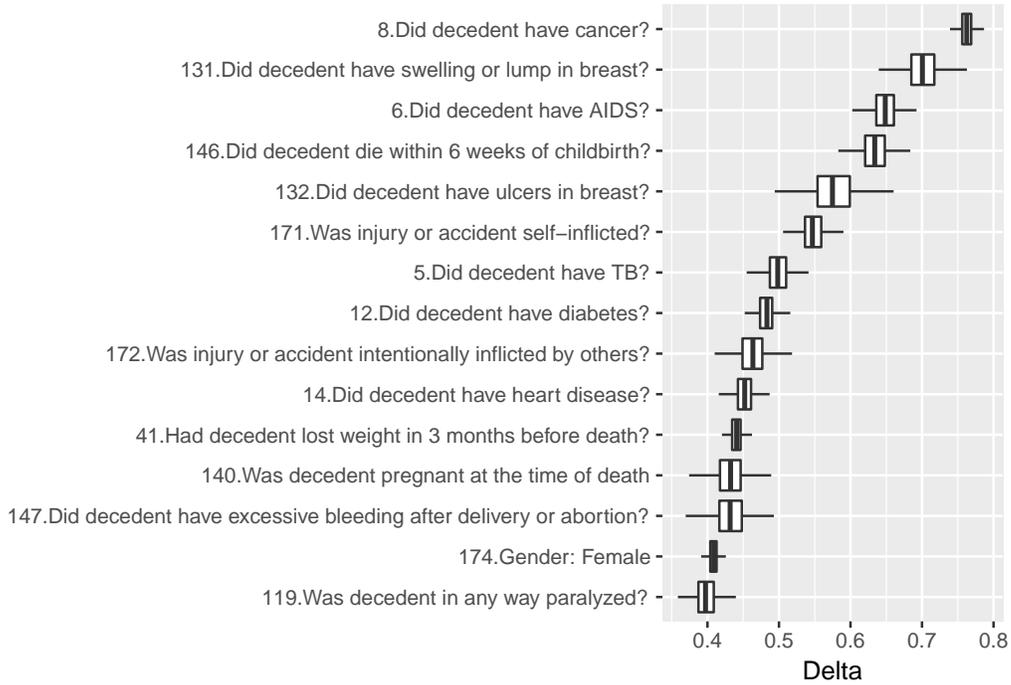}
\caption{Boxplot of $\delta$ for top 15 predictors in descending order of the posterior mean. The index of each predictor is in the supplementary materials.}
\label{fig:measure}
\end{center}
\end{figure}

We compare the proposed method with state of the art VA classification algorithms. In addition to the conditionally independent model, we employ methods currently used in practice: InSilicoVA \citep{McCormick16}, InterVA \citep{Byass12}, King-Lu \citep{King08, King10} and Tariff \citep{James11, Serina15}. These methods assign causes without the information about constellations or clusters of symptoms. King-Lu is the exception that takes into account dependence between symptoms but it can incorporate only a small subset of symptoms at a time. On the other hand, the proposed method can jointly model all predictors without the assumption of conditional independence. Details of the competitors are in the supplementary materials.

Figure \ref{fig:csmf} reports the boxplot of CSMF accuracy for each scenario. It shows the proposed method works well, producing the accuracy equal to or higher than the competitors in all scenarios. As for the competitors, the order changes in each scenario, and InterVA and Tariff often report small values. Although King-Lu takes into account dependence between symptoms, the performance can be lower than the other competitors with the assumption of conditional independence. This is probably because it relies on modeling of small subsets of predictors. All the methods show relatively low values in the scenario where the target site is Pemba Island with the largest gap of the distributions of causes between the target and training sites. Figure \ref{fig:correlation} shows the boxplot of correlation between the actual and predicted numbers of deaths per cause. The result is similar in that the proposed method works as well as or better than the competitors in all scenarios, and the values are relatively small for all the methods in the scenario with the target site Pemba Island. In addition, we conduced the sensitivity analysis of the prior and the number of random samples in (\ref{eq:mcmc-r}) for the proposed method and obtained stable results given in the supplementary materials.

Turning now to results for the measure of strength of association between the causes and predictors, we estimate $\delta$ in Section 2.2 using the proposed model. As in Figure 2, some items show high missing rates, and to obtain robust results, we include only predictors with a missing rate less than 5\%. Figure \ref{fig:measure} reports the boxplot of $\delta$ for top 15 predictors.  We observe that several items related to medical history show high dependence such as cancer and TB. There are also other types of items in the list. For example, predictors 140, 146, 147 are questions for women about pregnancy and childbirth, and predictors 171 and 172 are about injuries. The estimation result of $\delta$ for other predictors are in the supplementary materials. In addition, we compute the dependence measure in a cause-specific way by transforming the original variable with 34 causes into a binary variable taking 1 for a certain cause and 0 for the rest. The result indicates that important predictors vary with each cause. For example, a set of questions related to paralysis show relatively strong dependence with stroke, while the association of the medical history is much higher than other predictors in AIDS. These results are in the supplementary materials.

\section{Discussion} \label{sec:conclusion}

In the VA literature, there have been various statistical approaches for estimation of population distributions of causes of death. The key is how to model high dimensional binary predictors given a cause with many missing values. For simplicity, existing methods analyze survey data under strong assumptions such as conditional independence between symptoms. The contribution of this article is to analyze the PHMRC data using a new Bayesian method that flexibly captures complex interactions between symptoms without the restrictive assumption. In the proposed framework, one can measure strength of dependence of each symptom with the causes, which can be useful for the selection of questionnaire items in a future survey.

One future direction is to incorporate spatial information into the proposed model. Factors affecting cause of death vary through space depending on geographic characteristics, so two sites that are close to each other should largely share the same factors affecting cause of death. Therefore it may be more efficient to estimate distributions of deaths by cause by weighting more on neighboring areas. In addition the relationship between causes and questionnaire items may depend on space. Although this article assumes the conditional distribution of the symptoms given a cause is constant over space, one can extend it to $\pi(x\,|\,y,s)$ with spatial information $s$.

Another direction for future work is to generalize the proposed framework for survey weights.  To save cost and time many social surveys employ special data-collection designs such as stratified sampling that produce a biased sample. To adjust for a gap between the sample and the population, survey weights are constructed and distributed along with the data. When faced with data like that, it is necessary to incorporate the weights into statistical models for prediction.

\section*{Acknowledgment} 

This work was supported by The University of Washington eScience Moore/Sloan \& WRF Data Science Fellowship, JSPS Grant-in-Aid for Research Activity Start-up 16H06853, and grants K01HD078452 and 1R01HD086227 from the National Institute of Child Health and Human Development (NICHD). The results are generated mainly using Ox \citep{Doornik07}.

\bibliographystyle{chicago}
\bibliography{VA3,lancetVA,R01-VA}

\end{document}


\title
{Supplementary materials for ``Bayesian factor models for probabilistic cause of death assessment with verbal autopsies''}

\date{}

\maketitle

\vspace{-8mm}

\section{PHMRC data in the analysis}

Figures 1-6 report the barplot of causes of death for each study site and Tables 1-5 show the lists of causes of death and predictors in our analysis. We follow \citet{data} in the numbering of the causes. Following \citet{McCormick16}, the predictors are selected and transformed into dichotomous variables except that we use the female indicator as gender and skip the question for which no individual answered yes. 

\section{Difference in distributions of causes of death}

Figure 7 shows differences in empirical distributions of causes of death between test and training data in Section 4.

\section{Competitors}

The conditionally independent model predicts a cause of death by $\pi(y_i\,|\,x_i) \propto \prod_{j=1}^p \pi(x_{ij}\,|\,y_i)$ where beta(1,1) prior is assumed for $\pi(x_{ij}\,|\,y_i)$, leading to the posterior distribution, beta($1+\sum_{i=1}^n 1(x_{ij}=0,m_{ij}=0)$, $1+\sum_{i=1}^n 1(x_{ij}=1,m_{ij}=0)$) where $m_{ij}$ is the missing indicator taking 1 if $x_{ij}$ is missing and 0 if observed. The MCMC setting is the same as the proposed method in Section 3.

InterVA~\citep{Byass12} is one of the most popular automatic VA coding methods and has been extensively used in the practice. InterVA calculates the propensity of each cause given the symptoms that are present.  The original InterVA algorithm uses a list of physician-provided conditional probabilities of observing a symptom given each cause, in the form of rankings, without training data. It was later adapted to learn these conditional probabilities from training data in~\citet{McCormick16}. 

Tariff~\citep{James11} algorithm, as populated by the SmartVA-Analyze software, calculates the ranking of causes given observed symptoms using an additive score-based method. It should be noted that a modified algorithm, ``Tariff 2.0'', was proposed in~\citet{Serina15} with adjustments based on additional knowledge that is not publicly available. A discussion of the implementation was provided in the supplement materials of~\citet{McCormick16}. Unlike the methods that calculate individual cause assignments first and then aggregate to population distributions of causes, InSilicoVA~\citep{McCormick16} provides a Bayesian framework that assumes a generative model characterizing both CSMF at the population level, and the cause of death distributions at the individual level. 

All of the above methods assume the symptoms are conditionally independent given the underlying cause of death. King-Lu~\citep{King08, King10} is the only other method taking into account the dependence between symptoms. It estimates the population distribution of causes directly by constrained least square regressing on repeatedly sampled random subsets of symptoms.  

We estimate InterVA, Tariff and InSilicoVA using the R package, openVA~\citep{openvapkg, li2017openva}. For InterVA~\citep{intervapackage} and InSilicoVA~\citep{insilicovapkg}, the empirical conditional probabilities of observing a symptom given a cause are first transformed into ranks using the default InterVA interpretation table, as described in the supplement materials of~\citet{McCormick16}. For InSilicoVA, we ran the MCMC for $10,000$ iterations and saved every 10th iteration after discarding the first half of the chain. We used the default settings for Tariff~\citep{tariffpkg} and King-Lu~\citep{klpkg}. For InterVA, Tariff and King-Lu, we evaluate uncertainty in the estimates by the bootstrap with 100 random samples from training data.

\section{MCMC convergence}

To investigate posterior convergence, Figure \ref{fig:convergence} shows illustrative examples of sample paths and autocorrelations of the MCMC sample by the Bayesian factor model and the conditionally independent model in each scenario in Section 4. 

\section{Cross validation}

Figure \ref{fig:cv} shows the estimation result of 5-fold cross validation for the selection of an optimal number of latent factors in each scenario in Section 4.

\section{Sensitivity analysis}

To check the sensitivity of prior distributions, we estimated the distribution of causes of death in each scenario in Section 4 using Gamma($a$,$a$) prior for the precisions with $a=0.25,0.5,1,2,3$. Figure \ref{fig:prior} reports the boxplot of CSMF accuracy with these prior settings. With respect to the number of Monte Carlo simulation, $R$, we check the effects on the estimation result. Figure \ref{fig:R} reports the boxplot of CSMF accuracy with $R=100$, $200$, $300$, $400$, $500$ in each scenario in Section 4. 

\section{Association between causes of death and predictors}

Figure \ref{fig:delta} displays the boxplot of $\delta$ between 34 causes of death and each predictor except the ones with the missing rate more than 5\%. Figures \ref{fig:measure-stroke} and \ref{fig:measure-AIDS} show the boxplot of $\delta$ for top 15 predictors in stroke and AIDS in descending order of the posterior mean.

\begin{table}[h]
\centering
\caption{List of causes of death}
\begin{tabular}{cl|cl}
\hline
No.	& Cause of death & No. & Cause of death	 \\
\hline 
1 & AIDS & 18 & Leukemia/Lymphomas \\ 
2 & Asthma & 19 & Lung Cancer \\ 
3 & Bite of Venomous Animal & 20 & Malaria \\ 
4 & Breast Cancer & 21 & Maternal \\  
5 & Cervical Cancer & 22 & Other Cardiovascular Diseases \\ 
6 & Cirrhosis & 23 & Other Infectious Diseases \\
7 & Colorectal Cancer & 24 & Other Injuries \\ 
8 & COPD & 25 & Other Non-communicable Diseases \\ 
9 & Diabetes & 26 & Pneumonia \\ 
10 & Diarrhea/Dysentery & 27 & Poisonings \\
11 & Drowning & 28 & Prostate Cancer \\  
12 & Epilepsy & 29 & Renal Failure \\ 
13 & Esophageal Cancer & 30 & Road Traffic \\  
14 & Falls & 31 & Stomach Cancer \\  
15 & Fires & 32 & Stroke \\ 
16 & Homicide & 33 & Suicide \\  
17 & Acute Myocardial Infarction & 34 & TB \\  
\hline
\end{tabular}
\end{table}

\begin{figure}[h] 
\begin{center}
\includegraphics[scale=1]{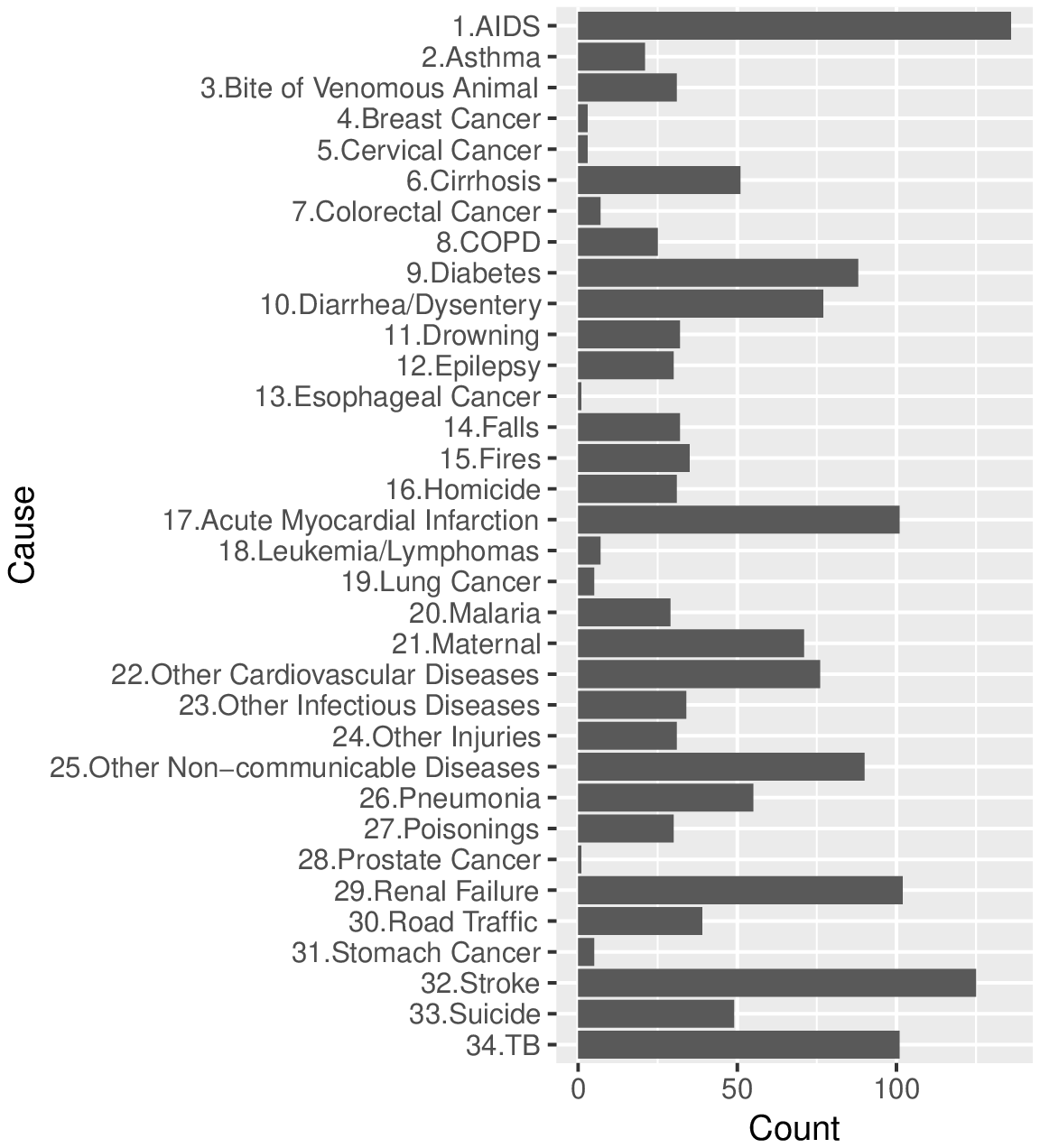}
\caption{Barplot of causes of death in Andhra Pradesh. We follow \citet{data} in the numbering of the causes.}
\end{center}
\end{figure}

\begin{figure}[h] 
\begin{center}
\includegraphics[scale=1]{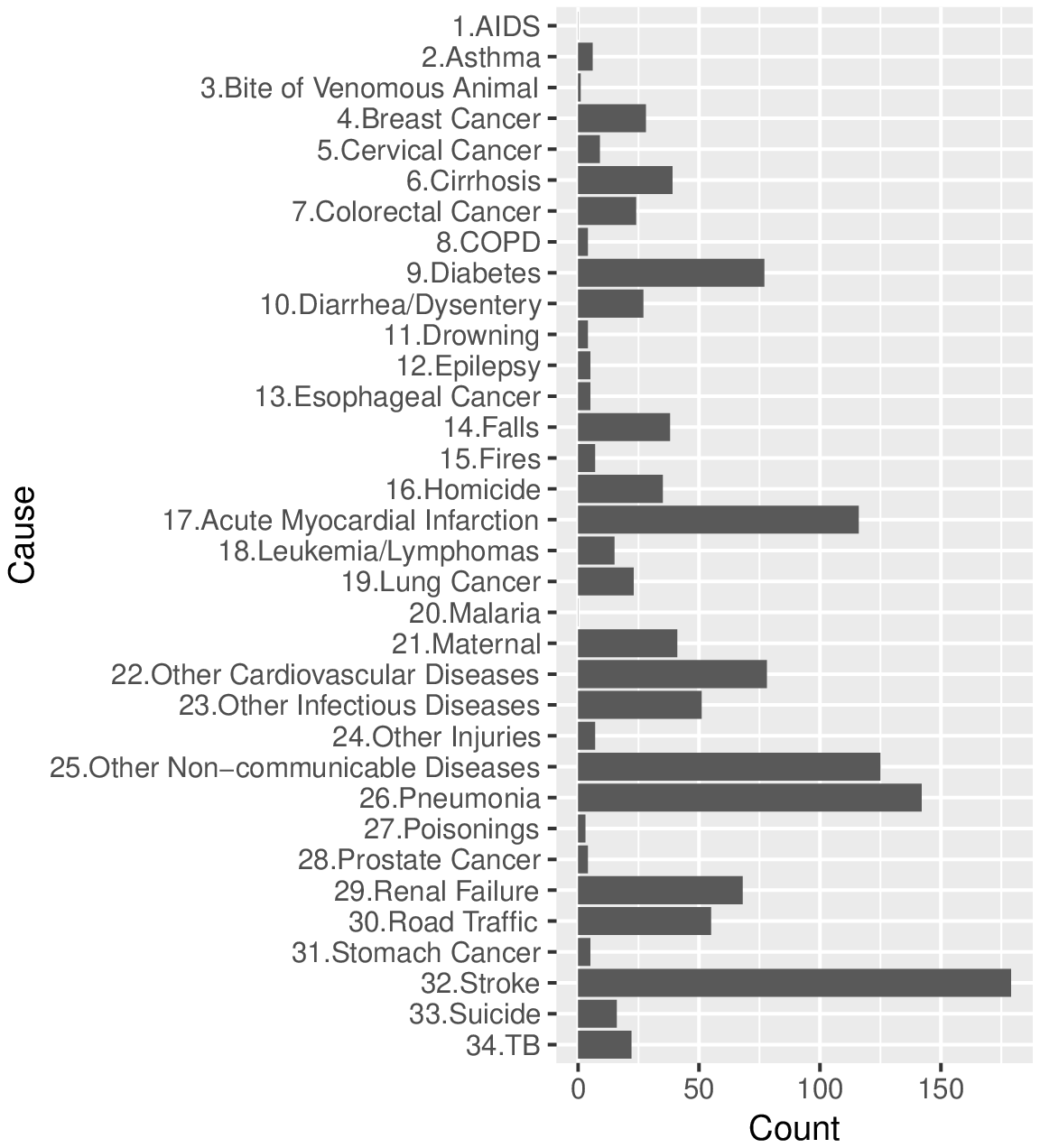}
\caption{Barplot of causes of death in Bohol. We follow \citet{data} in the numbering of the causes.}
\end{center}
\end{figure}

\begin{figure}[h] 
\begin{center}
\includegraphics[scale=1]{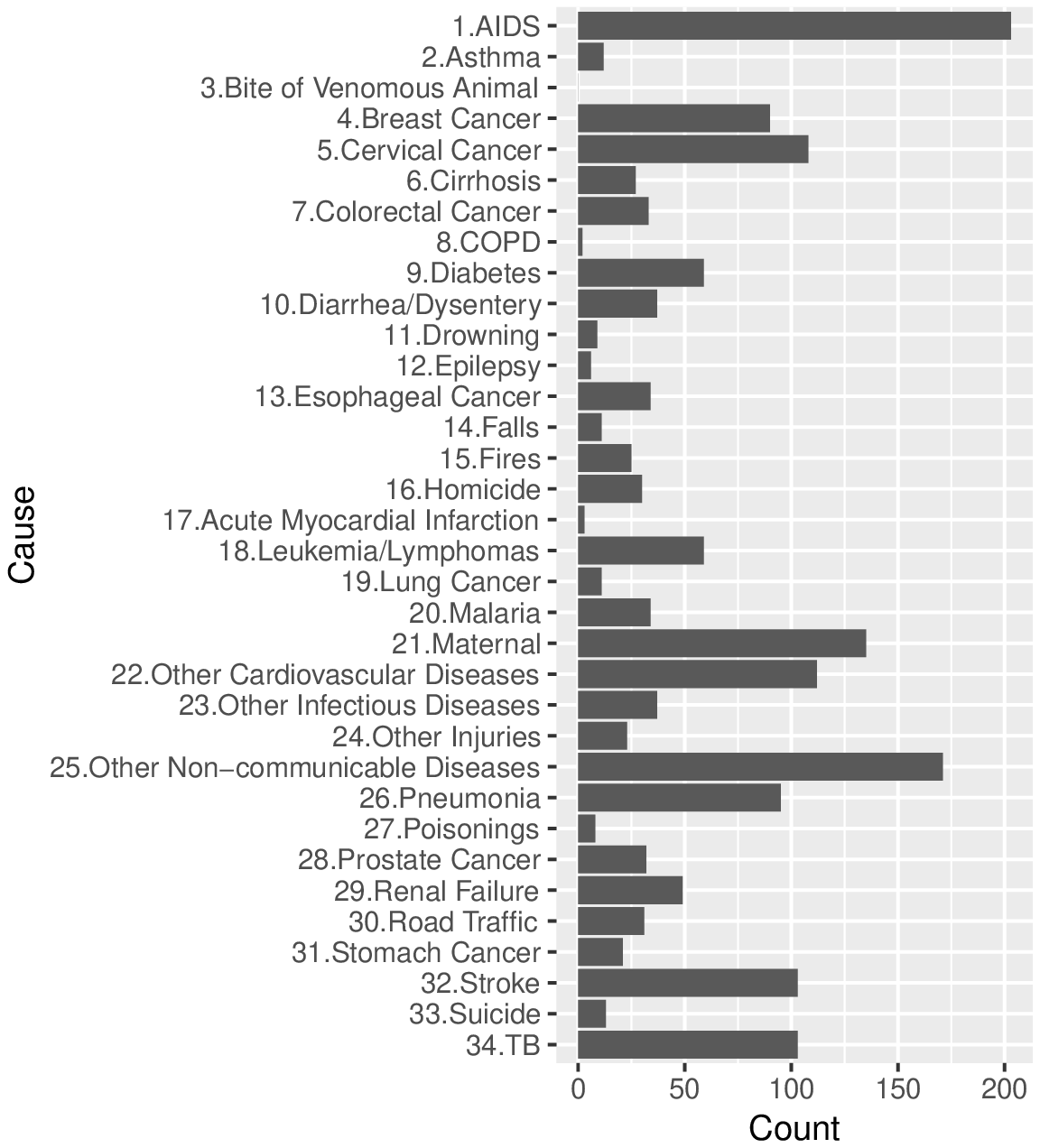}
\caption{Barplot of causes of death in Dar es Salaam. We follow \citet{data} in the numbering of the causes.}
\end{center}
\end{figure}

\begin{figure}[h] 
\begin{center}
\includegraphics[scale=1]{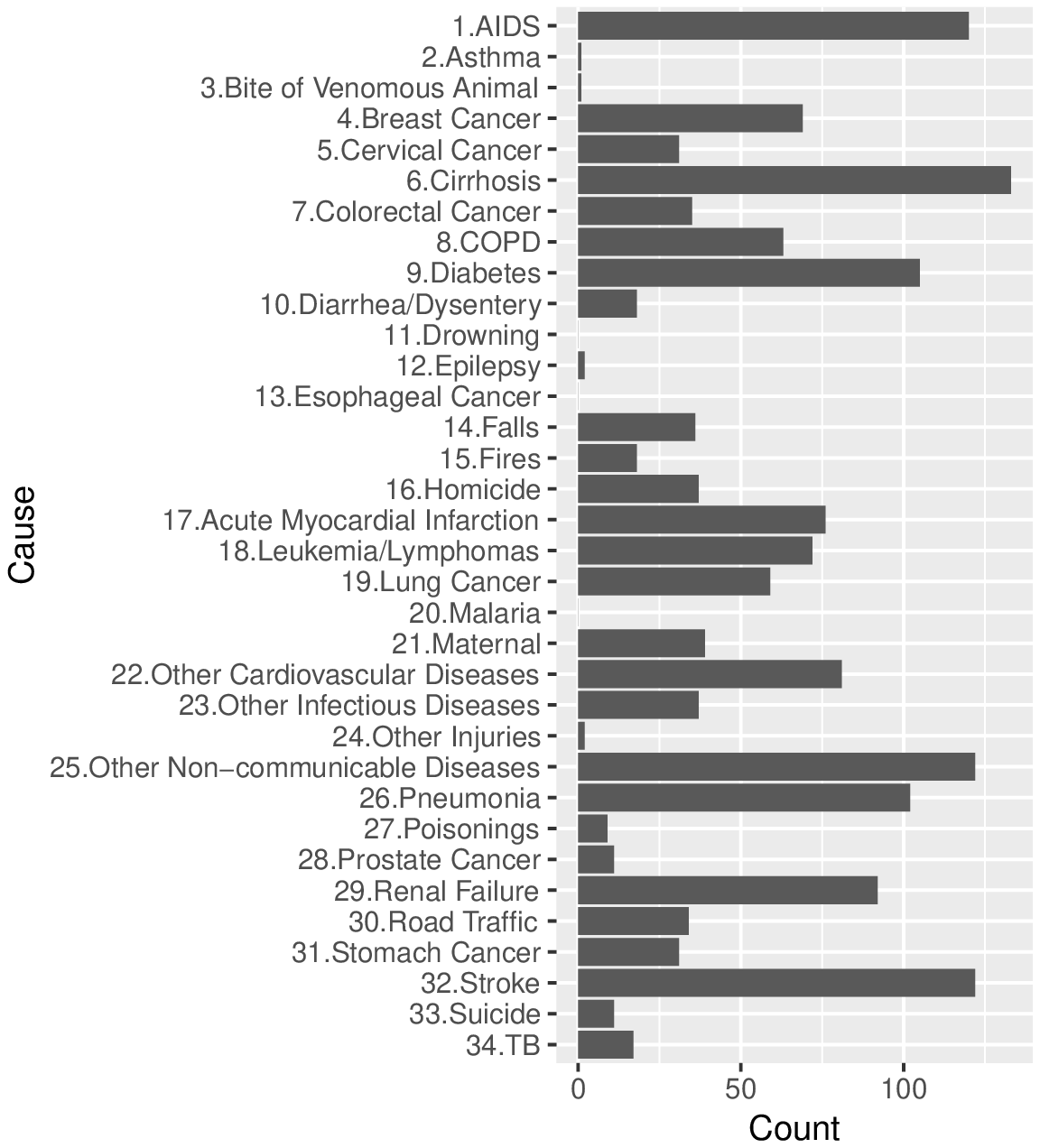}
\caption{Barplot of causes of death in Mexico City. We follow \citet{data} in the numbering of the causes.}
\end{center}
\end{figure}

\begin{figure}[h] 
\begin{center}
\includegraphics[scale=1]{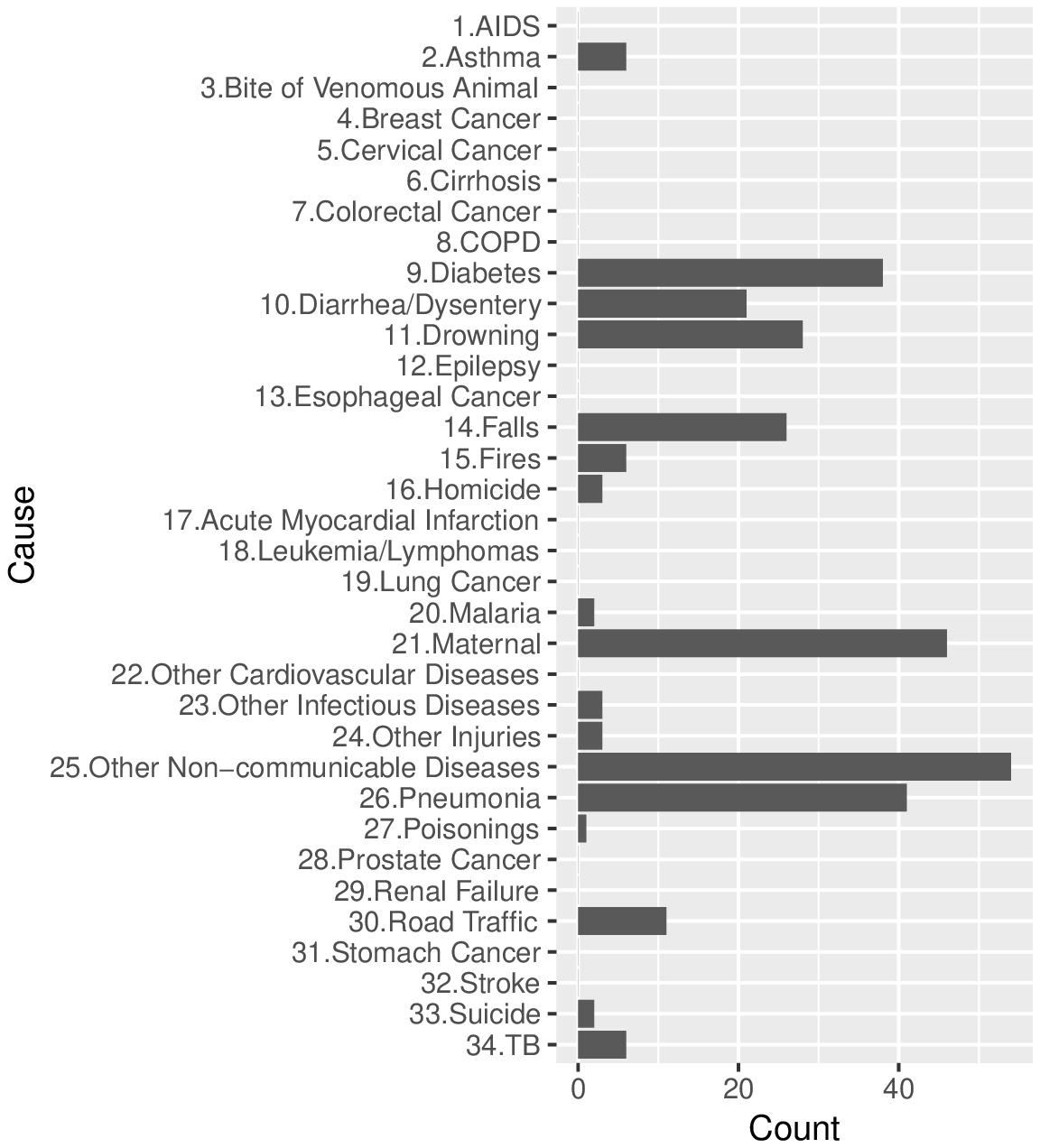}
\caption{Barplot of causes of death in Pemba Island. We follow \citet{data} in the numbering of the causes.}
\end{center}
\end{figure}

\begin{figure}[h] 
\begin{center}
\includegraphics[scale=1]{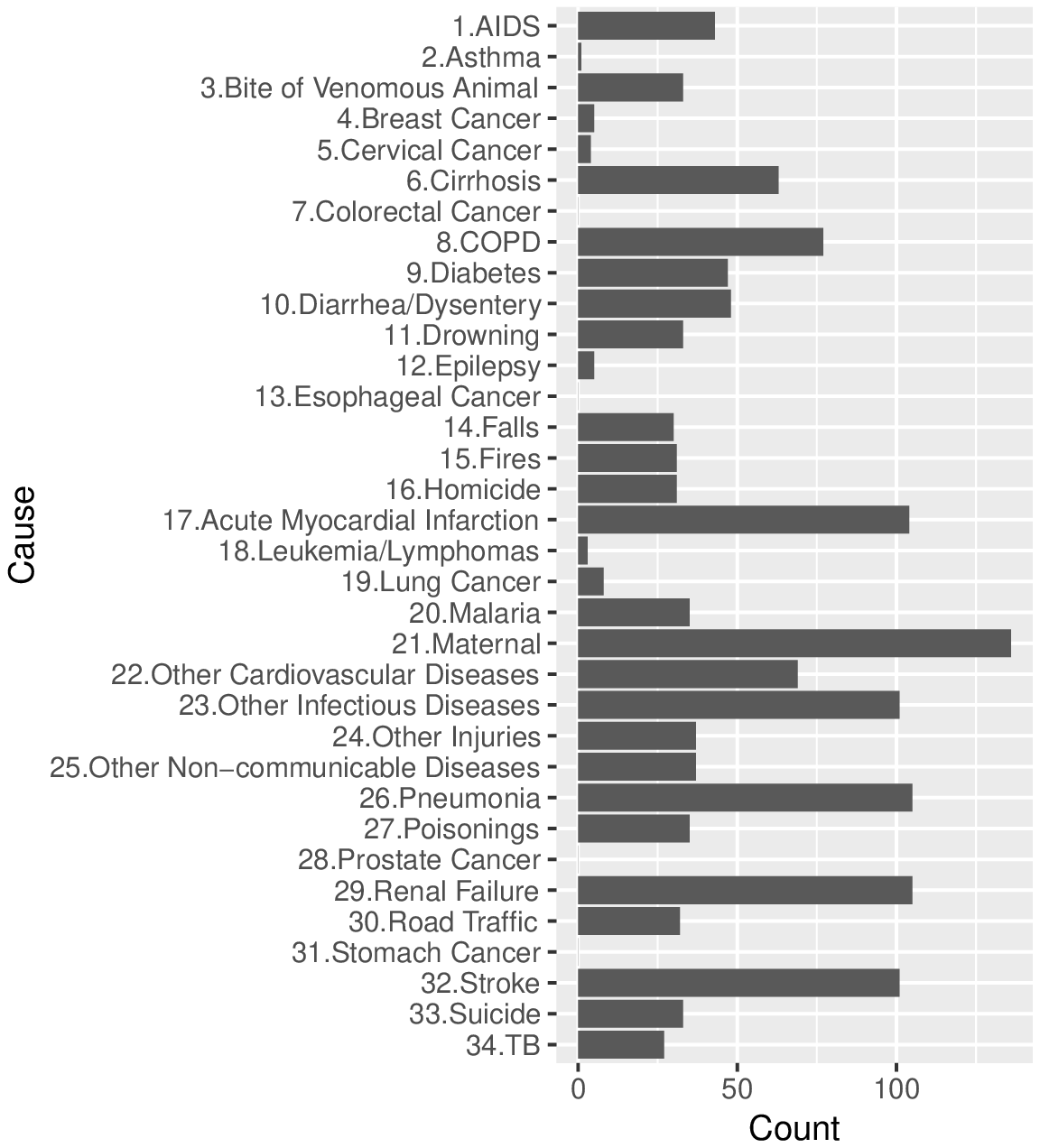}
\caption{Barplot of causes of death in Uttar Pradesh. We follow \citet{data} in the numbering of the causes.}
\end{center}
\end{figure}

\begin{table}[h]
\centering
\caption{List of predictors: 1-44}
\begin{tabular}{cl}
\hline
\multicolumn{1}{c}{No.}		& \multicolumn{1}{l}{Predictors}	 \\
\hline 
1 & Did decedent have asthma? \\
2 & Did decedent have hypertension?  \\
3 & Did decedent have obesity? \\ 
4 & Did decedent have stroke? \\ 
5 & Did decedent have TB? \\ 
6 & Did decedent have AIDS? \\ 
7 & Did decedent have arthritis? \\ 
8 & Did decedent have cancer? \\ 
9 & Did decedent have COPD? \\ 
10 & Did decedent have dementia? \\
11 & Did decedent have depression? \\ 
12 & Did decedent have diabetes? \\ 
13 & Did decedent have epilepsy? \\ 
14 & Did decedent have heart disease? \\ 
15 & For how long was decedent ill before s/he died? \\
16 & Did decedent have a fever? \\ 
17 & How many days did the fever last? \\ 
18 & Was there a moderate to severe fever? \\ 
19 & Was there a continuous fever? \\ 
20 & Was there an on and off fever? \\ 
21 & Did decedent have sweating with the fever? \\ 
22 & Did decedent have a rash? \\ 
23 & How many days did decedent have the rash? \\
24 & Was there a rash on the face? \\ 
25 & Was there a rash on the trunk? \\ 
26 & Was there a rash on the extremities? \\ 
27 & Was there a rash everywhere? \\
28 & Was there a rash in other locations? \\ 
29 & Was there a rash on the face? [second choice] \\ 
30 & Was there a rash on the trunk? [second choice] \\ 
31 & Was there a rash on the extremities? [second choice] \\ 
32 & Was there a rash in other locations? [second choice] \\ 
33 & Did decedent have sores? \\ 
34 & Did the sores have clear fluid or pus? \\ 
35 & Did decedent have itching of skin? \\
36 & Did decedent have an ulcer (pit) on the foot? \\ 
37 & Did the ulcer ooze pus? \\ 
38 & For how many days did the ulcer ooze pus? \\ 
39 & Did decedent experience pins and needles in their feet? \\ 
40 & Did decedent have blue lips? \\ 
41 & Had decedent lost weight in the three months prior to death? \\ 
42 & Was there moderate to large weight loss? \\ 
43 & Did decedent look pale? \\
44 & Did decedent have yellow discoloration of the eyes? \\
\hline
\end{tabular}
\end{table}

\begin{table}[h]
\centering
\caption{List of predictors: 45-88}
\begin{tabular}{cl}
\hline
\multicolumn{1}{c}{No.}		& \multicolumn{1}{l}{Predictors}	 \\
\hline 
45 & For how long did decedent have the yellow discoloration? \\
46 & Did decedent have ankle swelling? \\
47 & For how long did decedent have ankle swelling? \\ 
48 & Did decedent have puffiness of the face? \\ 
49 & For how long did decedent have puffiness of the face? \\ 
50 & Did decedent have general puffiness all over his/her body? \\ 
51 & For how long did decedent have puffiness all over his/her body? \\ 
52 & Did decedent have a lump in the neck? \\ 
53 & Did decedent have a lump in the armpit? \\ 
54 & Did decedent have a lump in the groin? \\
55 & Did decedent have a cough? \\ 
56 & For how long did decedent have a cough? \\ 
57 & Did the cough produce sputum? \\ 
58 & Did decedent cough blood? \\ 
59 & Did decedent have difficulty breathing? \\ 
60 & For how long did decedent have difficulty breathing? \\ 
61 & Was the breathing difficulty continuous? \\ 
62 & Was the breathing difficulty on and off? \\ 
63 & Did the breathing difficulty get worse in the lying position? \\ 
64 & Did the breathing difficulty get worse in the sitting position? \\ 
65 & Did the breathing difficulty get worse in the walking position? \\ 
66 & Did the breathing difficulty not get worse in any position? \\ 
67 & Did the breathing difficulty get worse in the lying position? [second choice] \\
68 & Did the breathing difficulty get worse in the sitting position? [second choice] \\ 
69 & Did the breathing difficulty get worse in the walking position? [second choice] \\ 
70 & Did the breathing difficulty not get worse in any position? [second choice] \\ 
71 & Did decedent have fast breathing? \\
72 & For how long did decedent have fast breathing? \\ 
73 & Did decedent wheeze? \\ 
74 & Did decedent experience pain in the chest in the month preceding death? \\ 
75 & Did the pain last more than 24 hours? \\ 
76 & Was the pain during physical activity? \\ 
77 & Was there pain located in the chest? \\ 
78 & Was there pain located in the left arm? \\ 
79 & Was there pain located in other places? \\
80 & Was there difficulty swallowing both solids and liquids? \\ 
81 & For how long before death did decedent have loose or liquid stools? \\ 
82 & Did decedent have a change in bowel habits? \\ 
83 & Was there blood in the stool? \\ 
84 & Was there blood in the stool up until death? \\ 
85 & Did decedent stop urinating? \\ 
86 & Did decedent vomit in the week preceding the death? \\ 
87 & For how long before death did decedent vomit? \\
88 & Was there blood in the vomit? \\
\hline
\end{tabular}
\end{table}

\begin{table}[h]
\centering
\caption{List of predictors: 89-132}
\begin{tabular}{cl}
\hline
\multicolumn{1}{c}{No.}		& \multicolumn{1}{l}{Predictors}	 \\
\hline 
89 & Was the vomit black? \\
90 & Did decedent have difficulty swallowing? \\
91 & For how long before death did decedent have difficulty swallowing? \\ 
92 & Was the difficulty with swallowing with solids, liquids, or both? \\ 
93 & Did decedent have pain upon swallowing? \\ 
94 & Did decedent have belly pain? \\ 
95 & For how long before death did decedent have belly pain? \\ 
96 & Was there pain in the lower belly? \\ 
97 & Was there pain in the upper belly? \\ 
98 & Did decedent have a more than usual protruding belly? \\
99 & For how long before death did decedent have a protruding belly? \\ 
100 & How rapidly did decedent develop the protruding belly? \\ 
101 & Did decedent have any mass in the belly? \\ 
102 & For how long before death did decedent have a mass in the belly? \\ 
103 & Did decedent have headaches? \\ 
104 & For how long before death did decedent have headaches? \\ 
105 & Was the onset of the headache fast? \\ 
106 & Did decedent have a stiff neck? \\ 
107 & For how long before death did decedent have stiff neck? \\ 
108 & Did decedent experience a period of loss of consciousness? \\ 
109 & Was there a sudden loss of consciousness? \\ 
110 & For how long did the period of loss of consciousness last? \\ 
111 & Did it continue until death? \\
112 & Did decedent experience a period of confusion at any time in the three months prior to death? \\ 
113 & For how long did the period of confusion last? \\ 
114 & Was there a sudden start to a period of confusion? \\ 
115 & Did decedent experience memory loss at any time in the three months prior to death? \\
116 & Did decedent have convulsions? \\ 
117 & For how long before death did the convulsions last? \\ 
118 & Did the person become unconscious immediately after the convulsions? \\ 
119 & Was decedent in any way paralyzed? \\ 
120 & For how long before death did decedent have paralysis? \\ 
121 & Paralyzed right side (arm and leg) \\ 
122 & Paralyzed other \\ 
123 & Paralyzed left side (arm and leg) \\
124 & Paralyzed lower part of body \\ 
125 & Paralyzed upper part of body \\ 
126 & Paralyzed one leg only \\ 
127 & Paralyzed one arm only \\ 
128 & Paralyzed whole body \\ 
129 & Paralyzed refused \\ 
130 & Paralyzed don't know \\ 
131 & Did decedent have any swelling or lump in the breast? \\
132 & Did decedent have any ulcers (pits) in the breast? \\
\hline
\end{tabular}
\end{table}

\begin{table}[h]
\centering
\caption{List of predictors: 133-175}
\begin{tabular}{cl}
\hline
\multicolumn{1}{c}{No.}		& \multicolumn{1}{l}{Predictors}	 \\
\hline 
133 & Had decedent periods stopped naturally because of menopause? \\
134 & Did decedent have vaginal bleeding after cessation of menstruation? \\
135 & Did decedent have vaginal bleeding other than her period? \\ 
136 & Was there excessive vaginal bleeding in the week prior to death? \\ 
137 & At the time of death was her period overdue? \\ 
138 & For how many weeks was her period overdue? \\ 
139 & Did decedent have a sharp pain in the belly shortly before death? \\ 
140 & Was decedent pregnant at the time of death? \\ 
141 & Did decedent die during an abortion? \\ 
142 & Did bleeding occur while she was pregnant? \\
143 & Did decedent have excessive bleeding during labor or delivery? \\ 
144 & Did decedent die during labor or delivery? \\ 
145 & Did decedent die within 6 weeks after having an abortion? \\ 
146 & Did decedent die within 6 weeks of childbirth? \\
147 & Did decedent have excessive bleeding after delivery or abortion? \\ 
148 & Did decedent have bad smelling vaginal discharge within 6 weeks after delivery or abortion? \\ 
149 & Did decedent use tobacco? \\ 
150 & Type of tobacco used: cigarettes \\ 
151 & Type of tobacco used: pipe \\ 
152 & Type of tobacco used: chewing \\ 
153 & Type of tobacco used: local tobacco \\ 
154 & Type of tobacco used: other \\ 
155 & Type of tobacco used: refused \\
156 & Type of tobacco used: don't know \\ 
157 & How much pipe/chewing tobacco did decedent use daily? \\ 
158 & How many cigarettes did decedent smoke daily?  \\
159 & Did decedent drink alcohol? \\
160 & Did decedent drink moderate-high amounts of alcohol? \\ 
161 & Did decedent drink low amounts of alcohol? \\ 
162 & Did decedent suffer road traffic injury? \\ 
163 & Did decedent suffer fall? \\ 
164 & Did decedent suffer drowning? \\ 
165 & Did decedent suffer poisoning? \\ 
166 & Did decedent suffer bite/sting? \\ 
167 & Did decedent suffer burn? \\
168 & Was decedent a victim of violence? \\ 
169 & Decedent did not suffer any injuries \\ 
170 & Did decedent suffer other injury? \\ 
171 & Was the injury or accident self-inflicted? \\ 
172 & Was the injury or accident intentionally inflicted by someone else? \\ 
173 & How long did decedent survive after the injury? \\ 
174 & Gender: female \\ 
175 & Age \\
\hline
\end{tabular}
\end{table}


\begin{figure}[htpb] 
\begin{center}
\includegraphics[width=13cm]{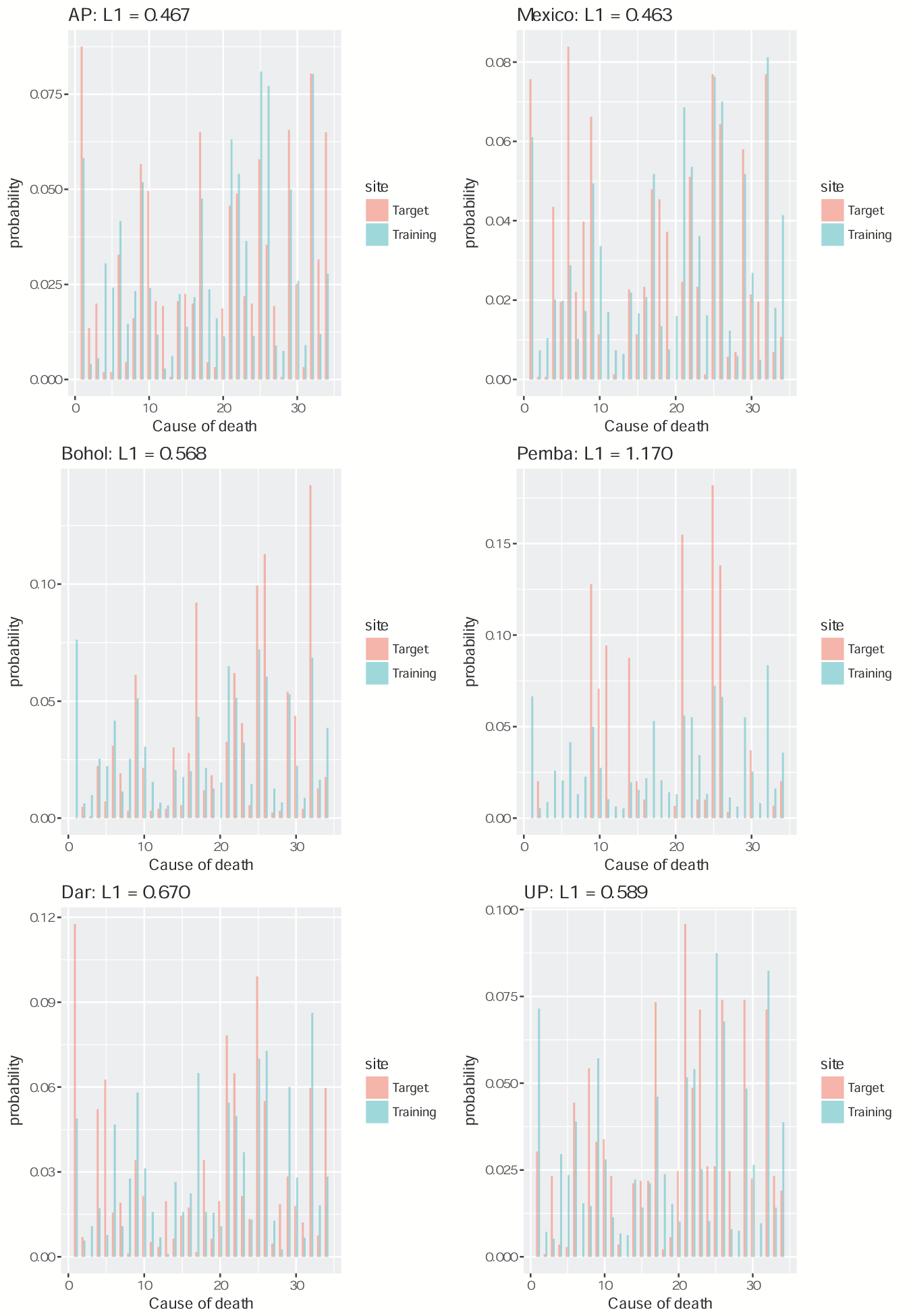}
\caption{Empirical distributions of causes of death in test and training data in Section 4. The left side shows the scenarios that the target sites are Andhra Pradesh (AP), Bohol and Dar es Salaam (Dar). The right side corresponds to Mexico City (Mexico), Pemba Island (Pemba) and Uttar Pradesh (UP). At the top of each figure, L1 indicates the $L_1$ distance between the two distributions.}
\label{fig:plot-pi}
\end{center}
\end{figure}


\begin{figure}[htpb] 
\begin{center}
\includegraphics[width=13.5cm,height=7cm]{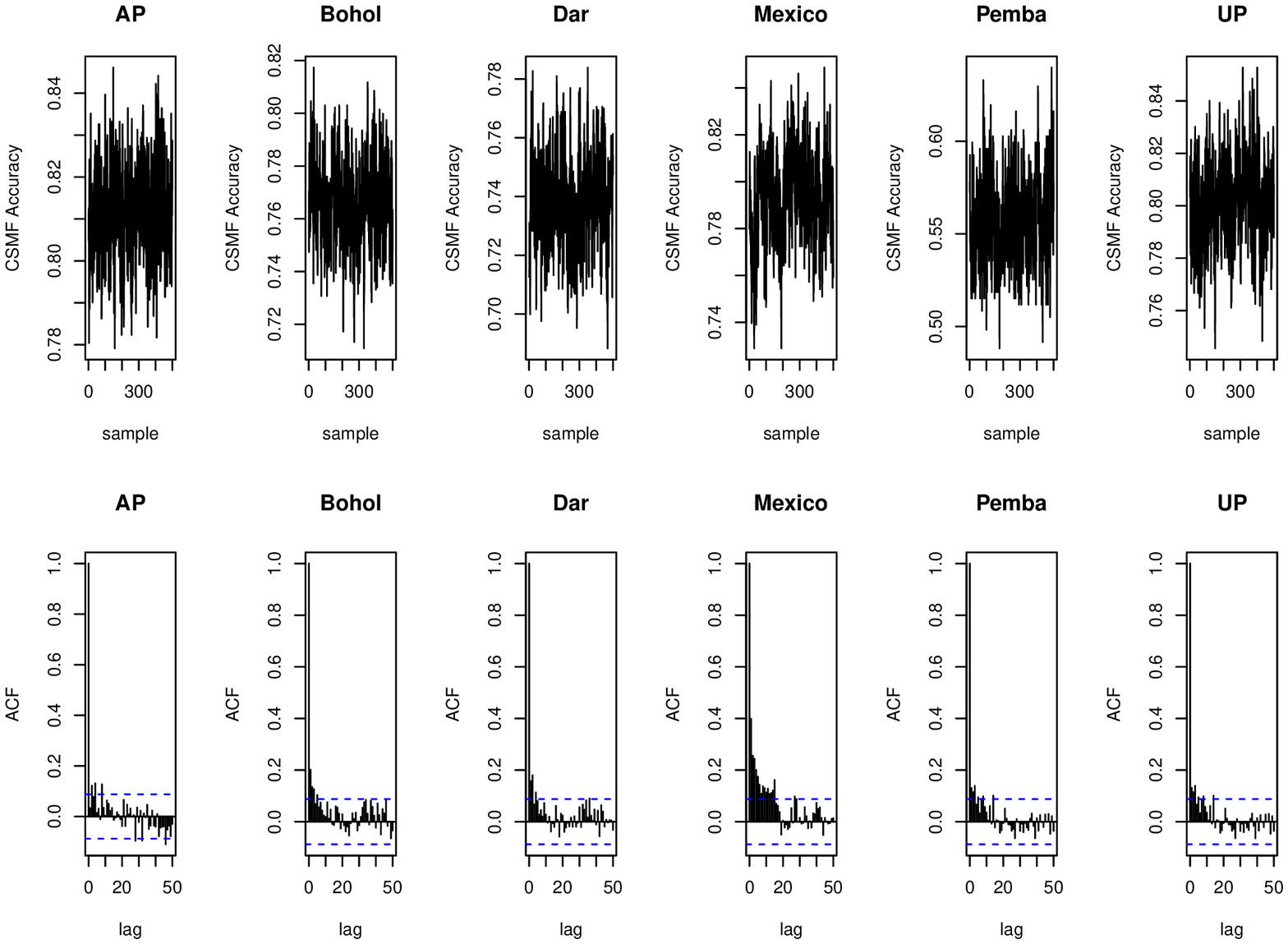}
\includegraphics[width=13.5cm,height=7cm]{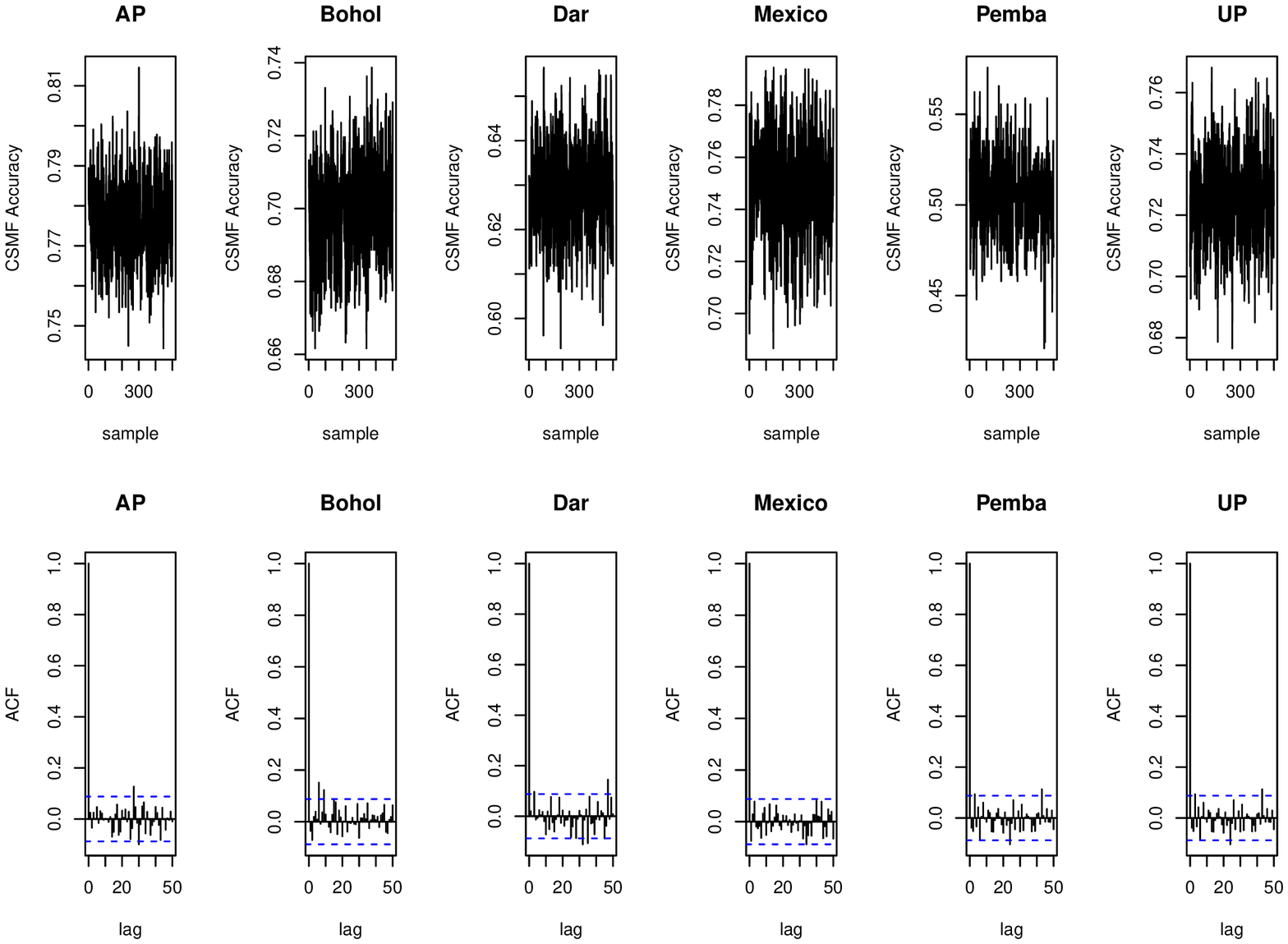}
\caption{Sample paths and autocorrelations of the MCMC sample for the Bayesian factor model (upper half) and the conditionally independent model (lower half). AP, Bohol, Dar, Mexico, Pemba and UP indicate the scenarios where the target sites are Andhra Pradesh, Bohol, Dar es Salaam, Mexico City, Pemba Island and Uttar Pradesh respectively.}
\label{fig:convergence}
\end{center}
\end{figure}


\begin{figure}[htbp]
  \begin{center}
    \begin{tabular}{c}
      \begin{minipage}{0.5\hsize}
        \begin{center}
          \includegraphics[scale=0.6]{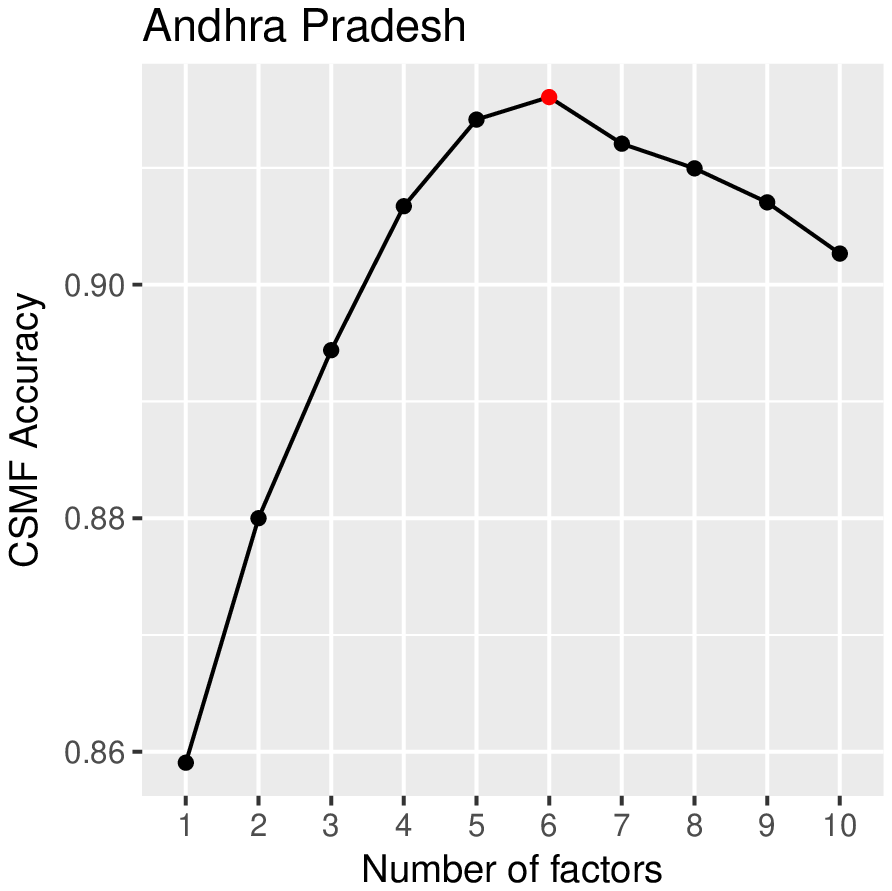}
        \end{center}
      \end{minipage}
      \begin{minipage}{0.5\hsize}
        \begin{center}
          \includegraphics[scale=0.6]{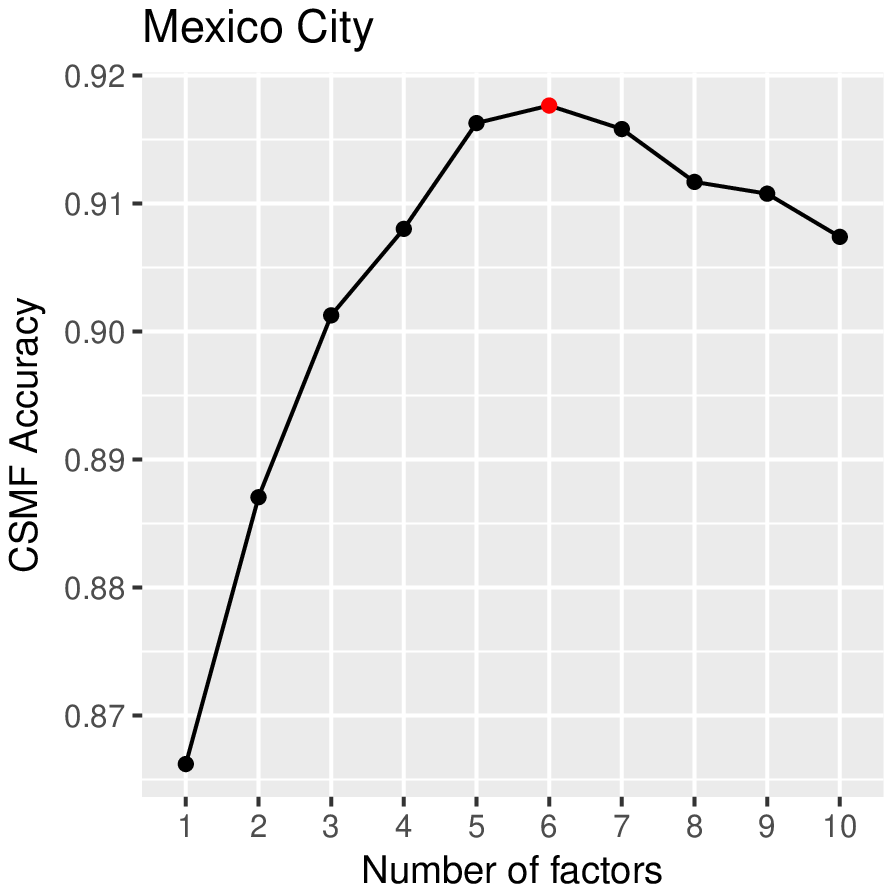}
        \end{center}
      \end{minipage}
    \end{tabular}
    \begin{tabular}{c}
      \begin{minipage}{0.5\hsize}
        \begin{center}
          \includegraphics[scale=0.6]{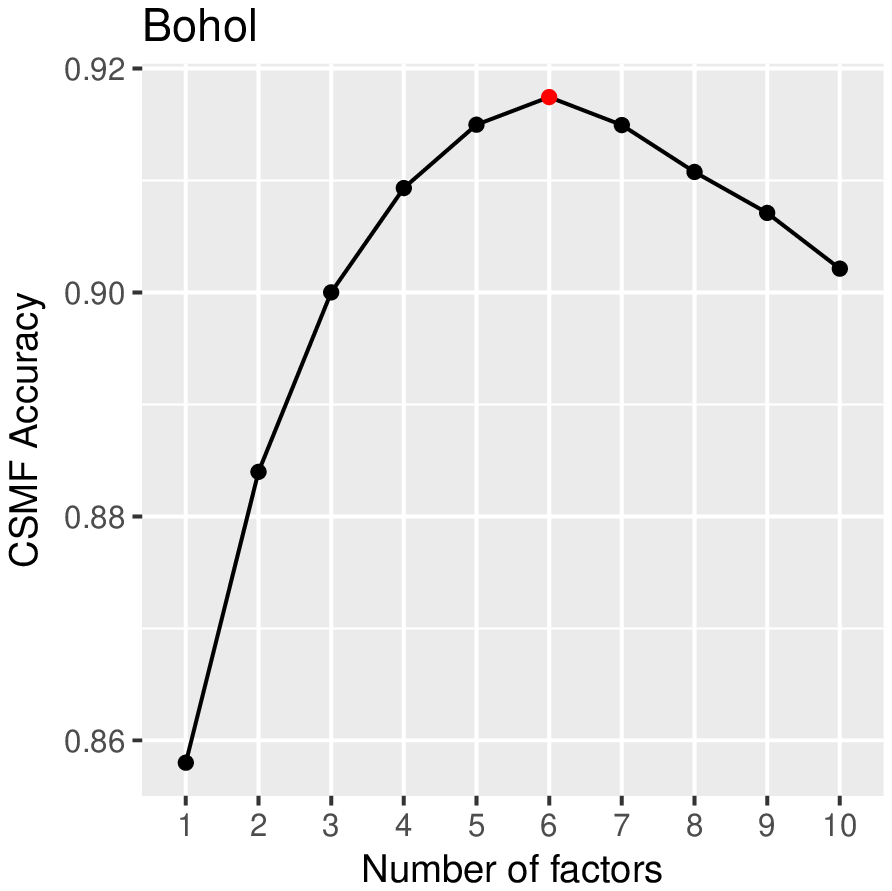}
        \end{center}
      \end{minipage}
      \begin{minipage}{0.5\hsize}
        \begin{center}
          \includegraphics[scale=0.6]{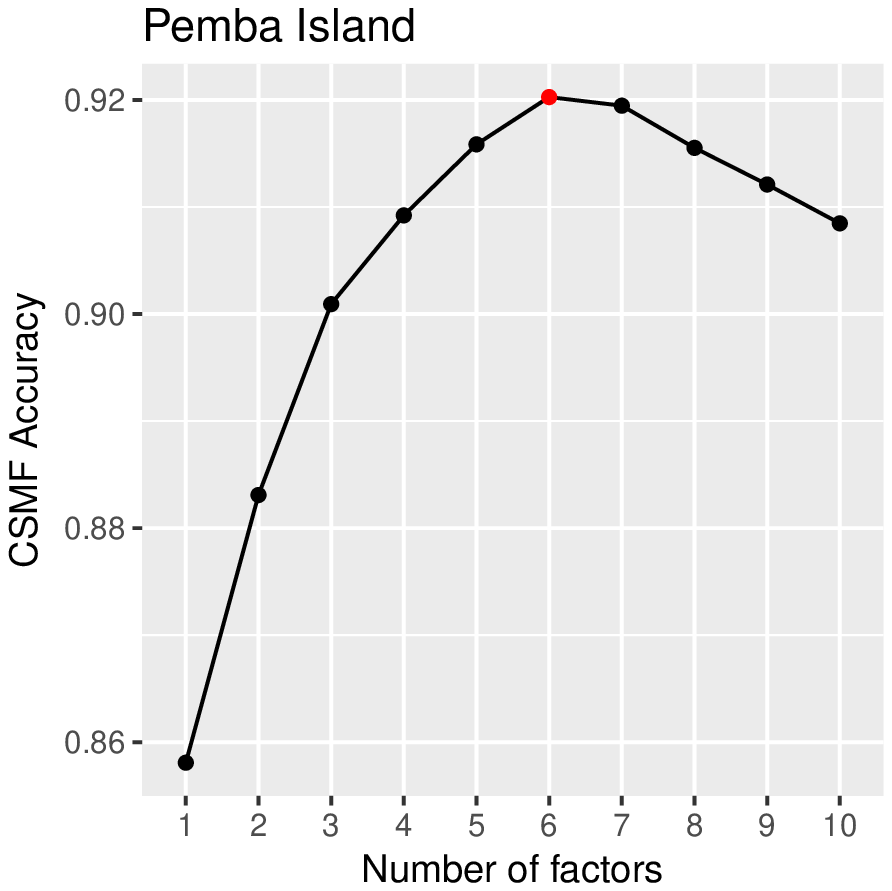}
        \end{center}
      \end{minipage}
    \end{tabular}
    \begin{tabular}{c}
      \begin{minipage}{0.5\hsize}
        \begin{center}
          \includegraphics[scale=0.6]{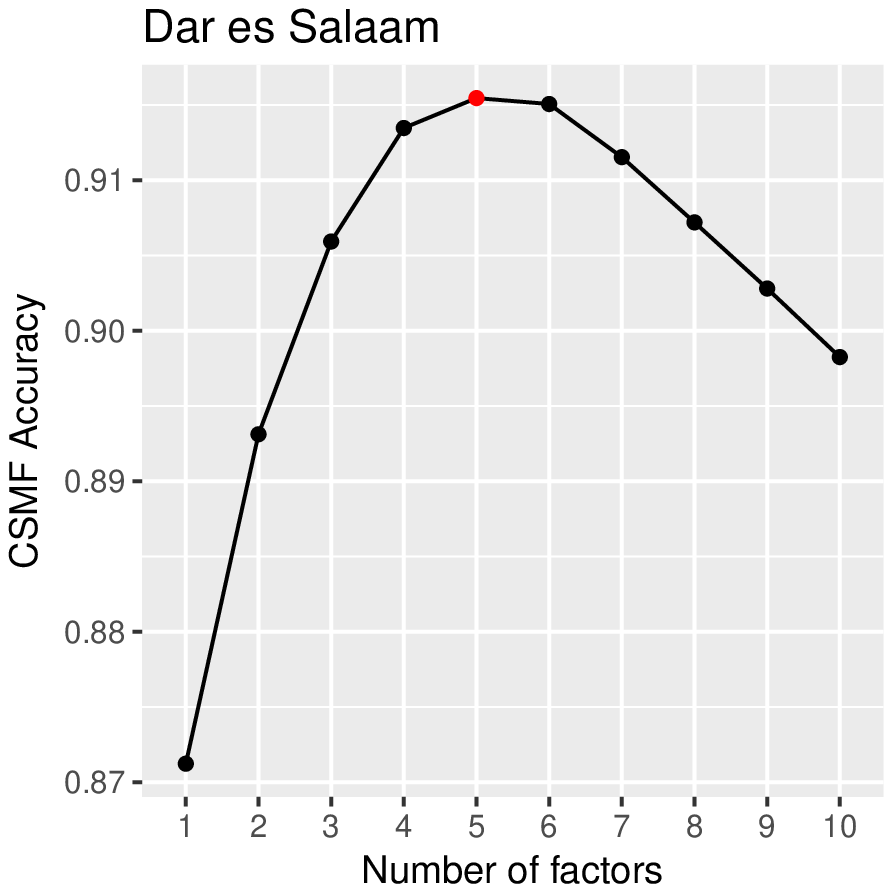}
        \end{center}
      \end{minipage}
      \begin{minipage}{0.5\hsize}
        \begin{center}
          \includegraphics[scale=0.6]{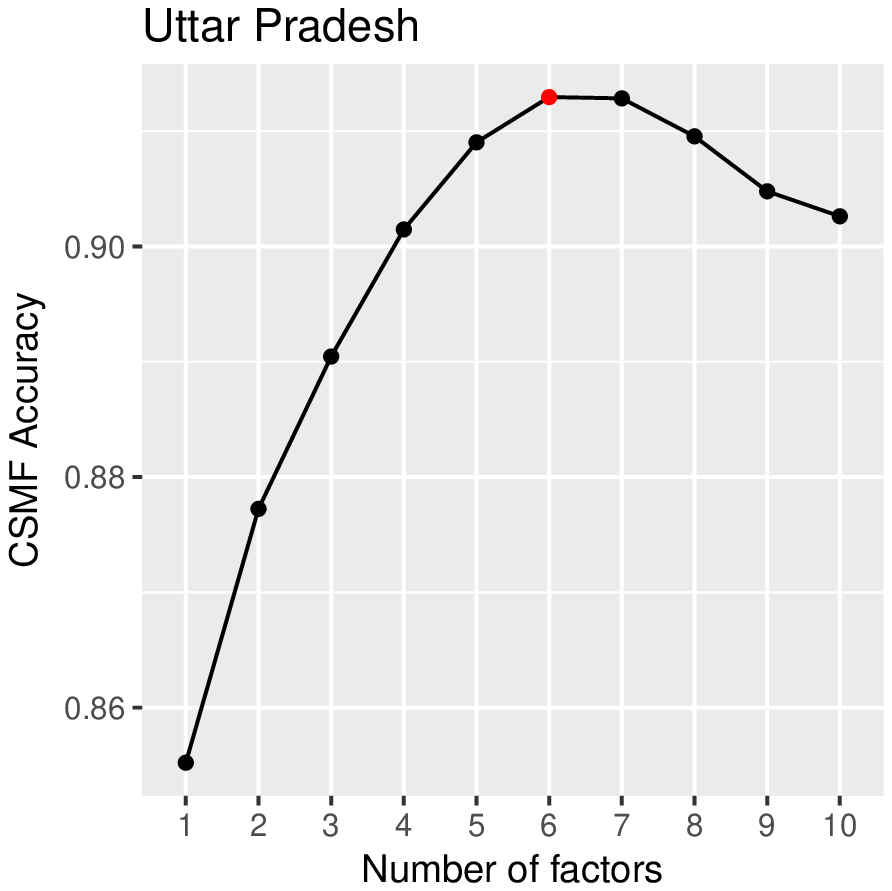}
        \end{center}
      \end{minipage}
    \end{tabular}
\caption{Estimation result of 5-fold cross validation in the scenarios where the target sites are Andhra Pradesh (top left), Bohol (middle left), Dar es Salaam (bottom left), Mexico City (top right), Pemba Island (middle right) and Uttar Pradesh (bottom right). The horizontal and vertical axes represent the number of factors and the posterior mean of CSMF accuracy. The red circle indicates the point with the highest accuracy.}
\label{fig:cv}
  \end{center}
\end{figure}


\begin{figure}[htbp]
  \begin{center}
    \begin{tabular}{c}
      \begin{minipage}{0.5\hsize}
        \begin{center}
          \includegraphics[scale=0.6]{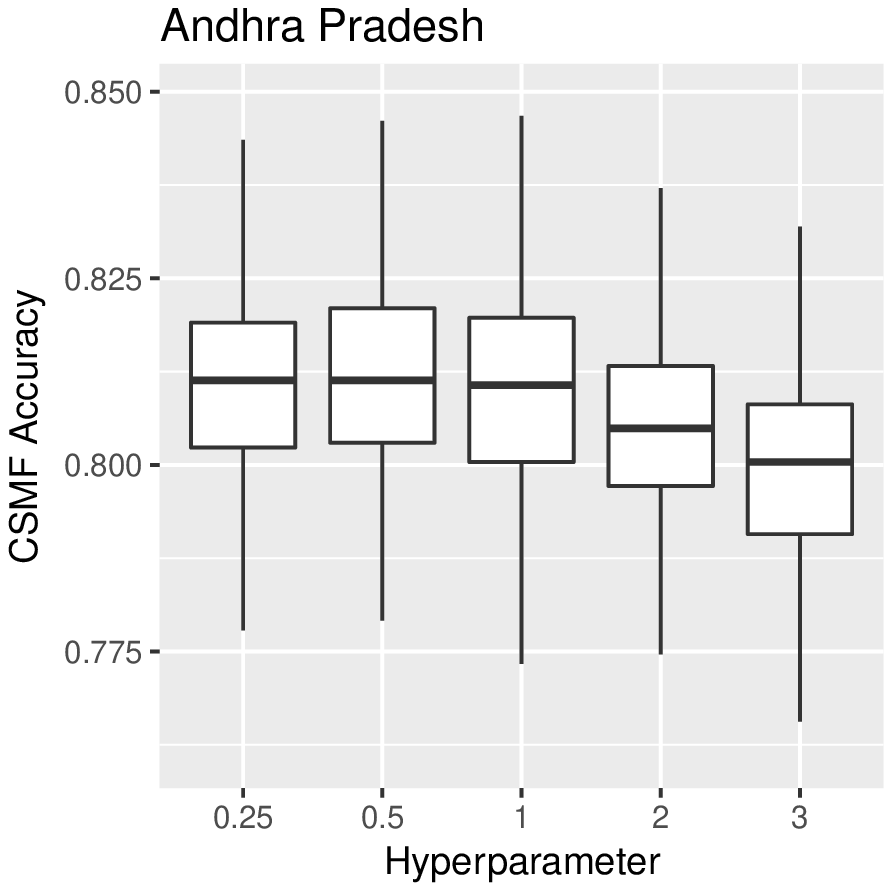}
        \end{center}
      \end{minipage}
      \begin{minipage}{0.5\hsize}
        \begin{center}
          \includegraphics[scale=0.6]{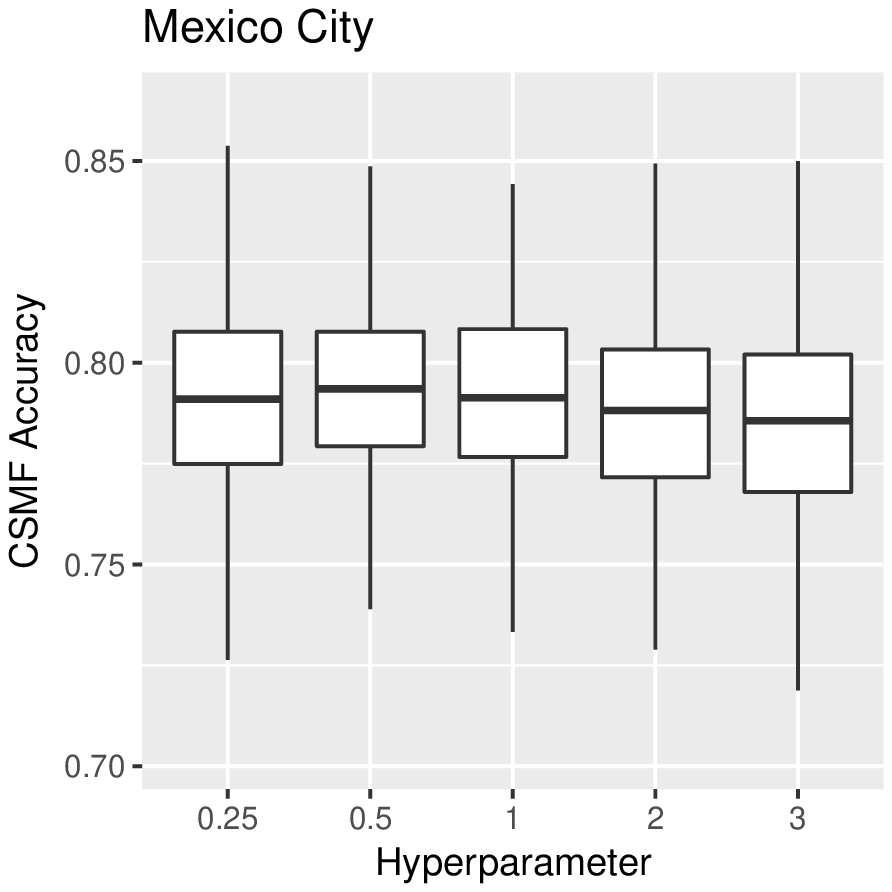}
        \end{center}
      \end{minipage}
    \end{tabular}
    \begin{tabular}{c}
      \begin{minipage}{0.5\hsize}
        \begin{center}
          \includegraphics[scale=0.6]{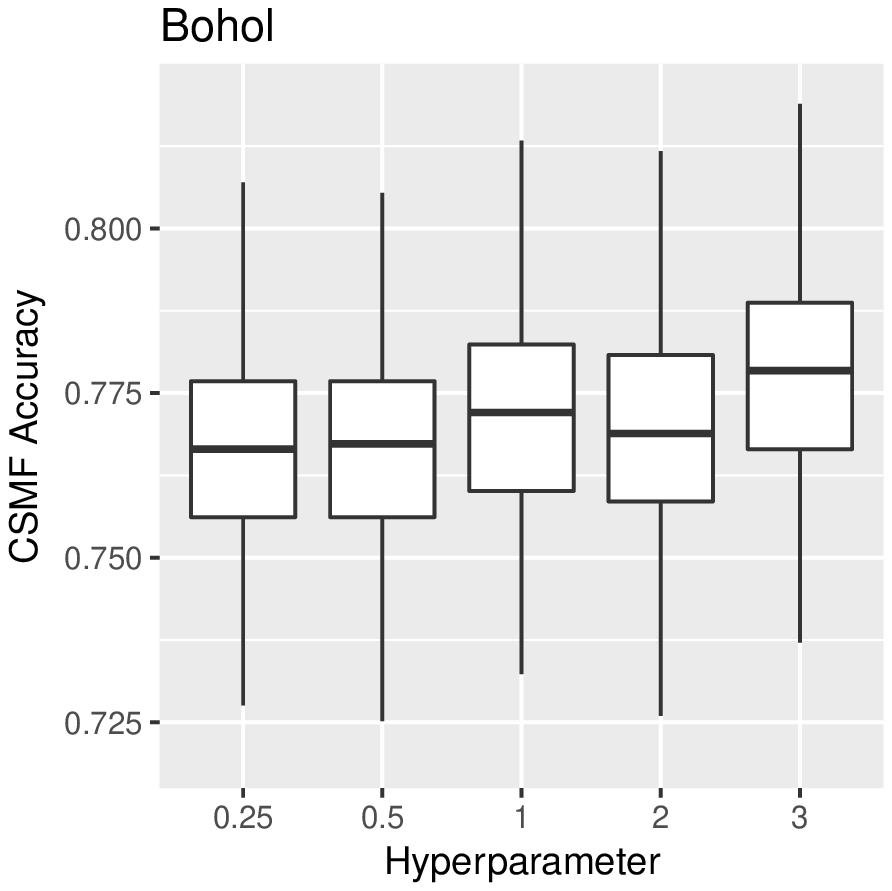}
        \end{center}
      \end{minipage}
      \begin{minipage}{0.5\hsize}
        \begin{center}
          \includegraphics[scale=0.6]{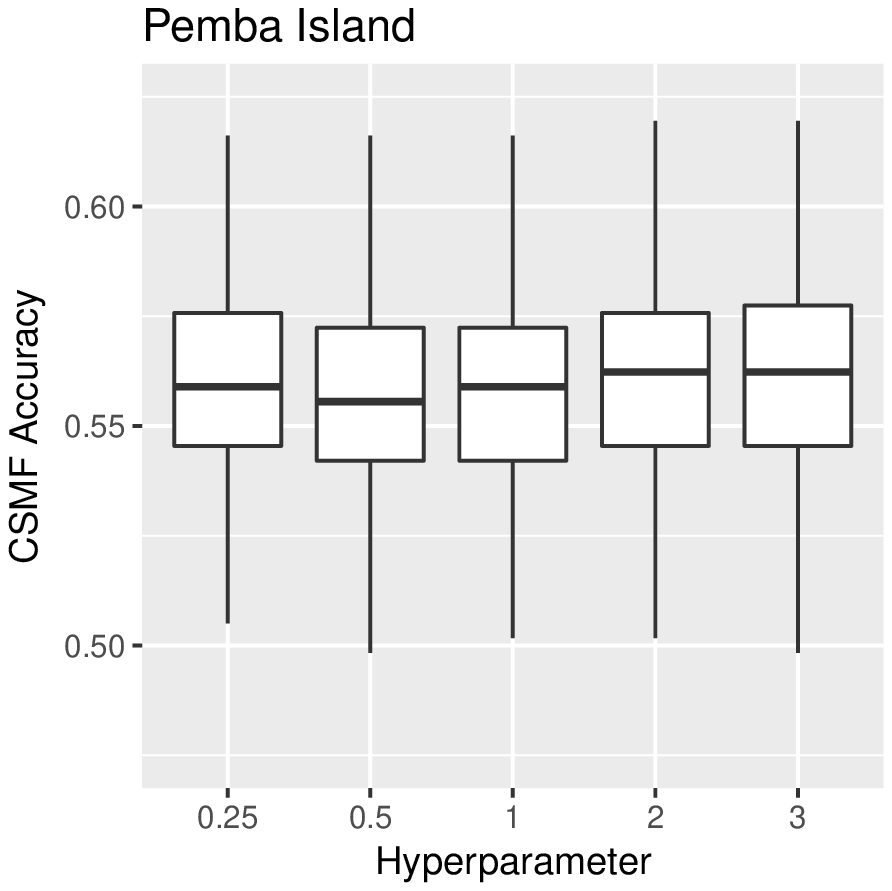}
        \end{center}
      \end{minipage}
    \end{tabular}
    \begin{tabular}{c}
      \begin{minipage}{0.5\hsize}
        \begin{center}
          \includegraphics[scale=0.6]{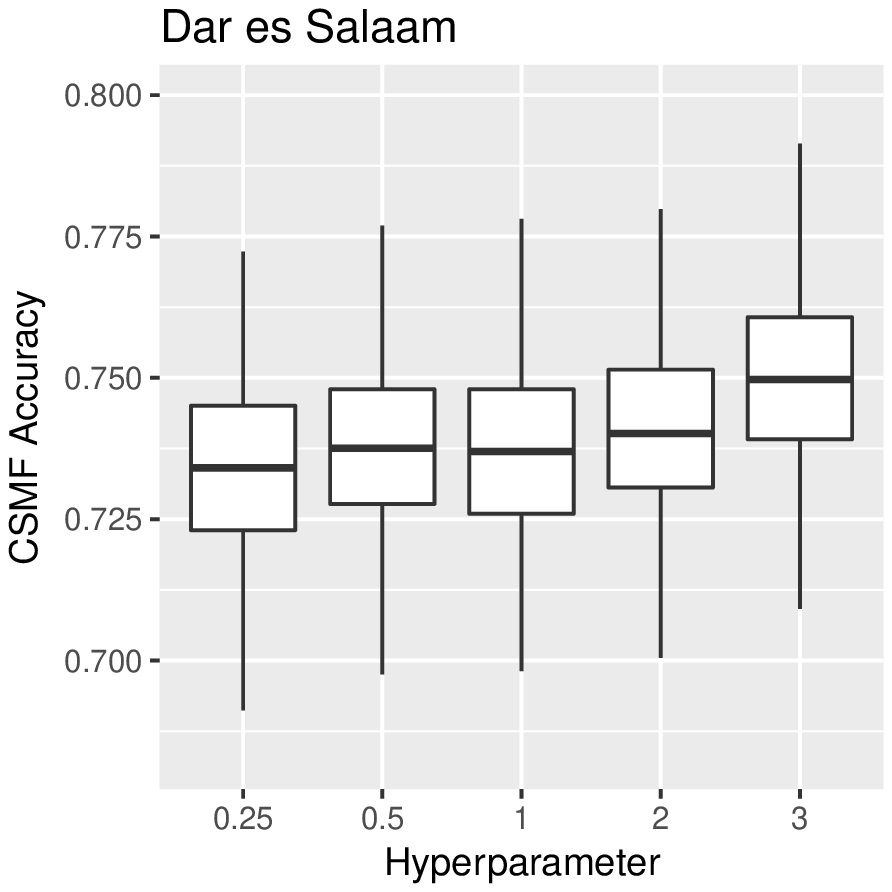}
        \end{center}
      \end{minipage}
      \begin{minipage}{0.5\hsize}
        \begin{center}
          \includegraphics[scale=0.6]{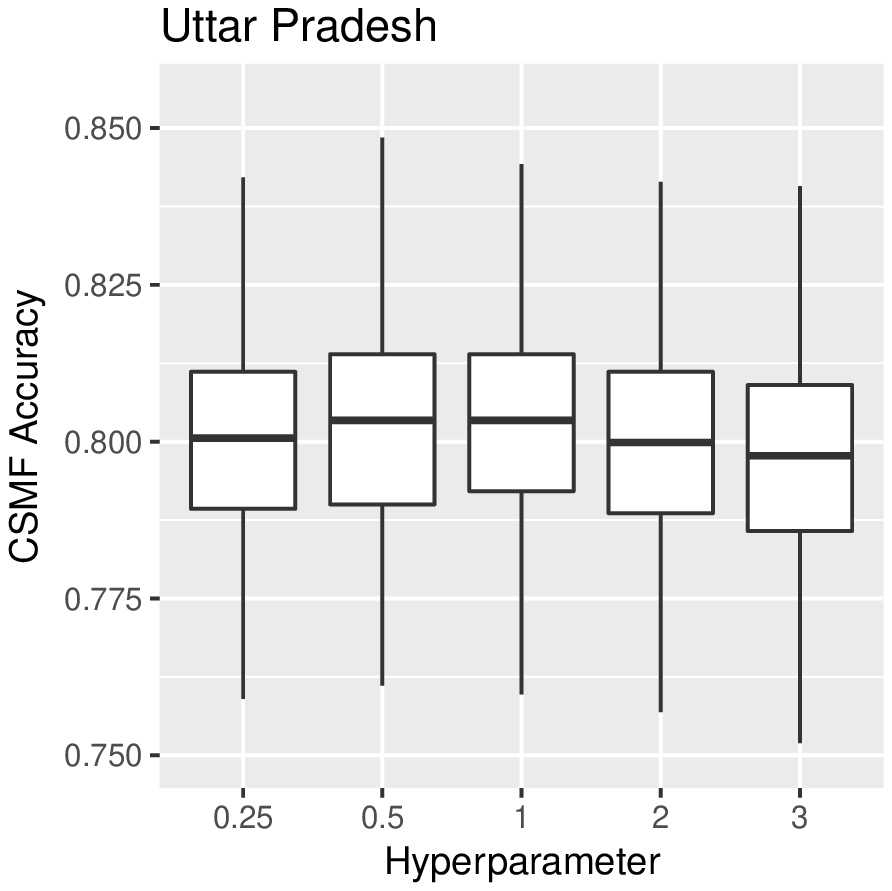}
        \end{center}
      \end{minipage}
    \end{tabular}
\caption{Boxplot of CSMF accuracy with Gamma($a$,$a$) prior for the precisions with $a=0.25$, 0.5, 1, 2, 3 in the scenarios where the target sites are Andhra Pradesh (top left), Bohol (middle left), Dar es Salaam (bottom left), Mexico City (top right), Pemba Island (middle right) and Uttar Pradesh (bottom right). The horizontal and vertical axes show the hyperparameter $a$ and CSMF accuracy.}
\label{fig:prior}
  \end{center}
\end{figure}

\begin{figure}[htbp]
  \begin{center}
    \begin{tabular}{c}
      \begin{minipage}{0.5\hsize}
        \begin{center}
          \includegraphics[scale=0.6]{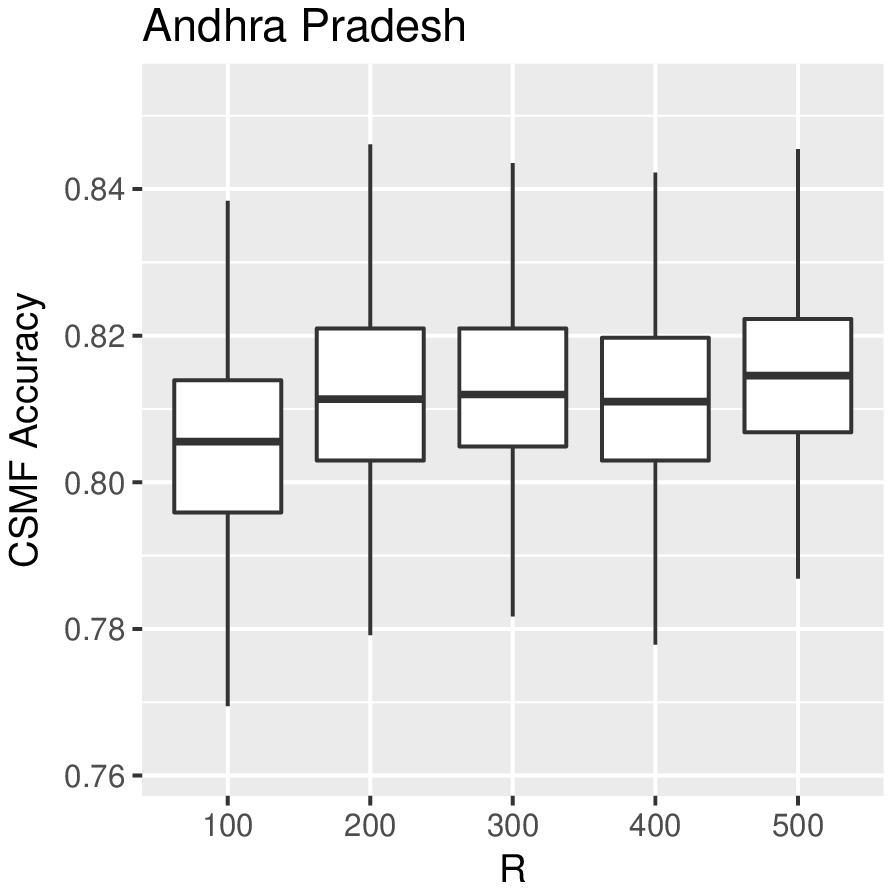}
        \end{center}
      \end{minipage}
      \begin{minipage}{0.5\hsize}
        \begin{center}
          \includegraphics[scale=0.6]{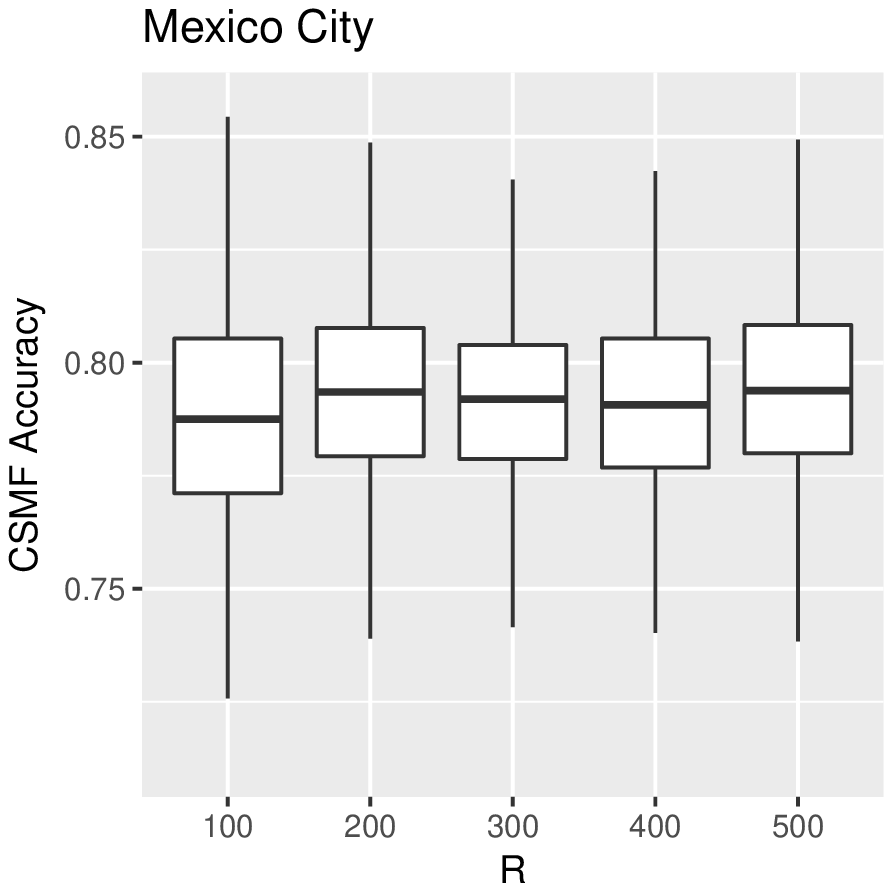}
        \end{center}
      \end{minipage}
    \end{tabular}
    \begin{tabular}{c}
      \begin{minipage}{0.5\hsize}
        \begin{center}
          \includegraphics[scale=0.6]{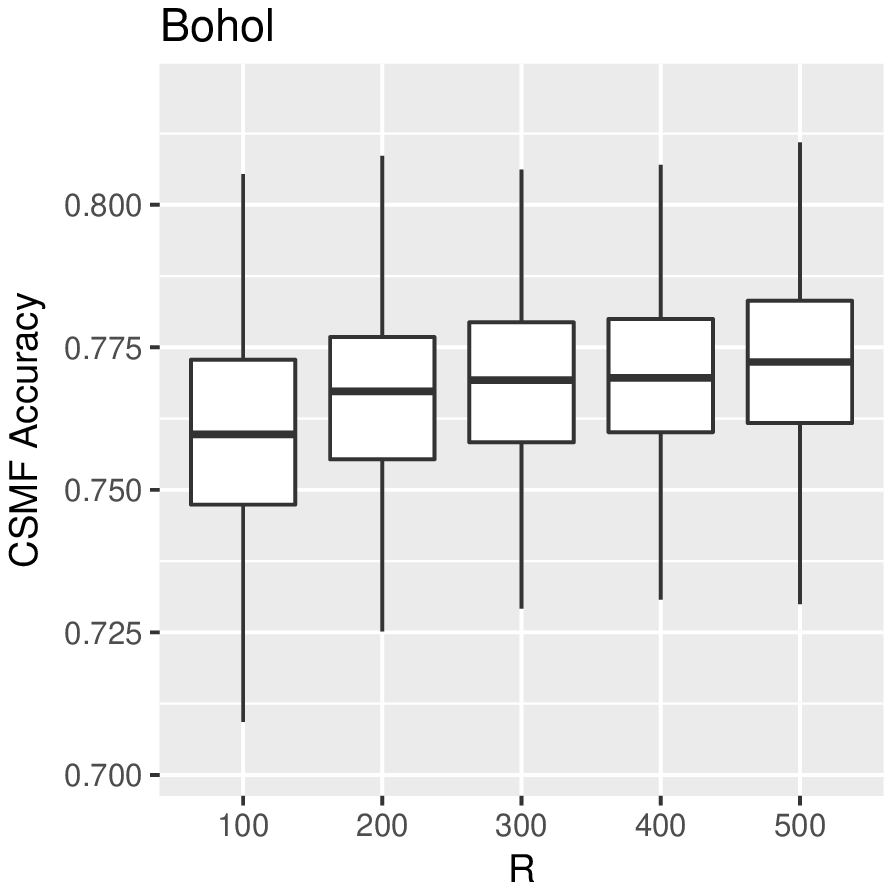}
        \end{center}
      \end{minipage}
      \begin{minipage}{0.5\hsize}
        \begin{center}
          \includegraphics[scale=0.6]{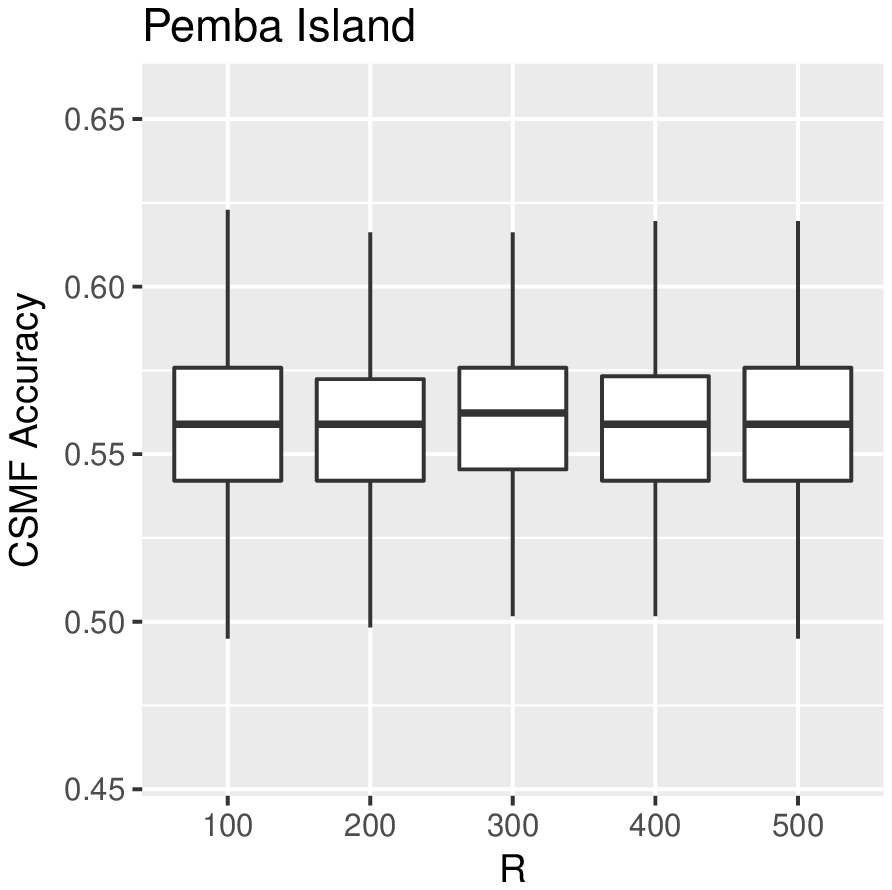}
        \end{center}
      \end{minipage}
    \end{tabular}
    \begin{tabular}{c}
      \begin{minipage}{0.5\hsize}
        \begin{center}
          \includegraphics[scale=0.6]{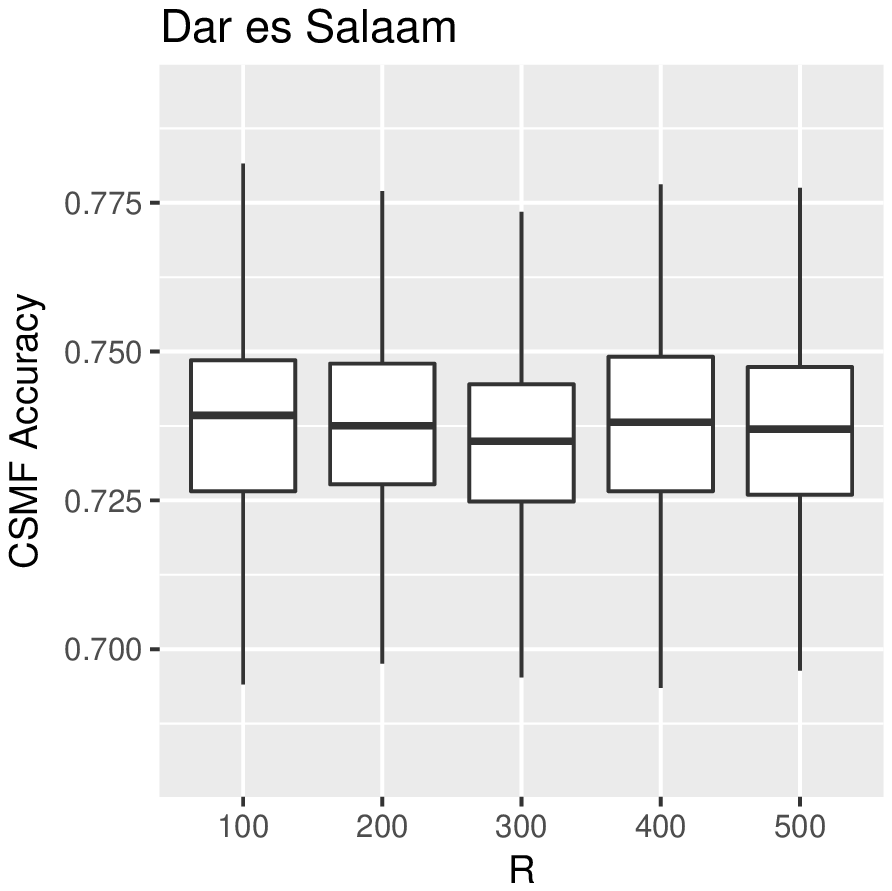}
        \end{center}
      \end{minipage}
      \begin{minipage}{0.5\hsize}
        \begin{center}
          \includegraphics[scale=0.6]{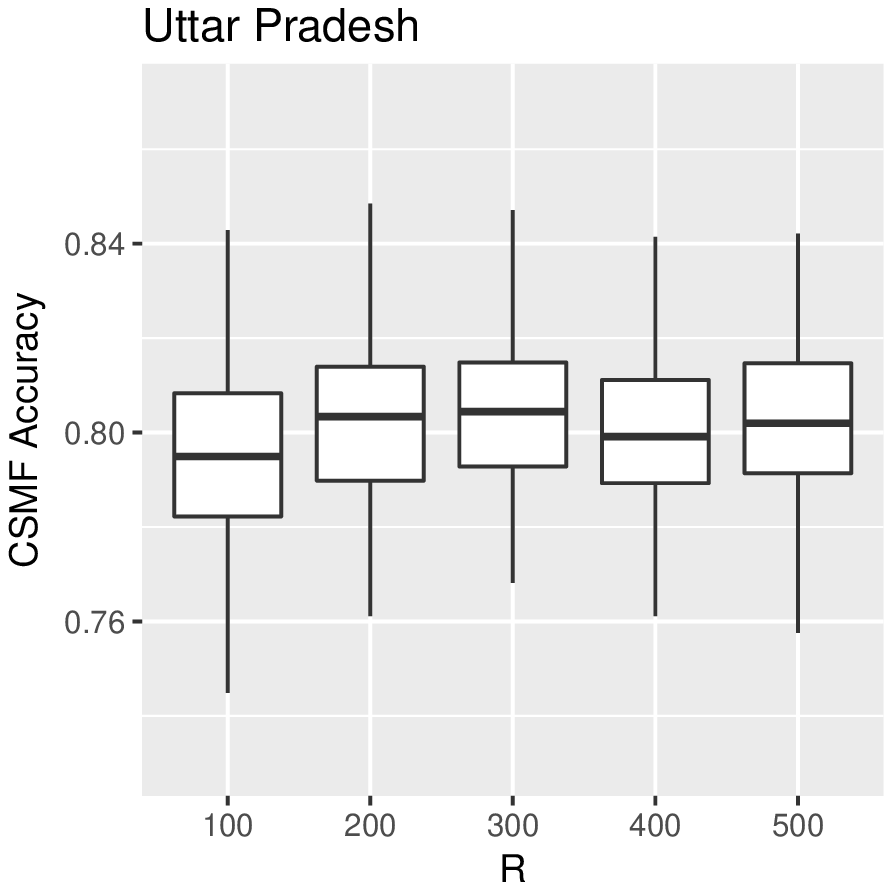}
        \end{center}
      \end{minipage}
    \end{tabular}
\caption{Boxplot of CSMF accuracy with $R=100$, 200, 300, 400, 500 in the scenarios where the target sites are Andhra Pradesh (top left), Bohol (middle left), Dar es Salaam (bottom left), Mexico City (top right), Pemba Island (middle right) and Uttar Pradesh (bottom right). The horizontal and vertical axes represent the number of the Monte Carlo simulation and CSMF accuracy.}
\label{fig:R}
  \end{center}
\end{figure}


\begin{figure}[htpb] 
\begin{center}
\includegraphics[width=13cm,height=6.5cm]{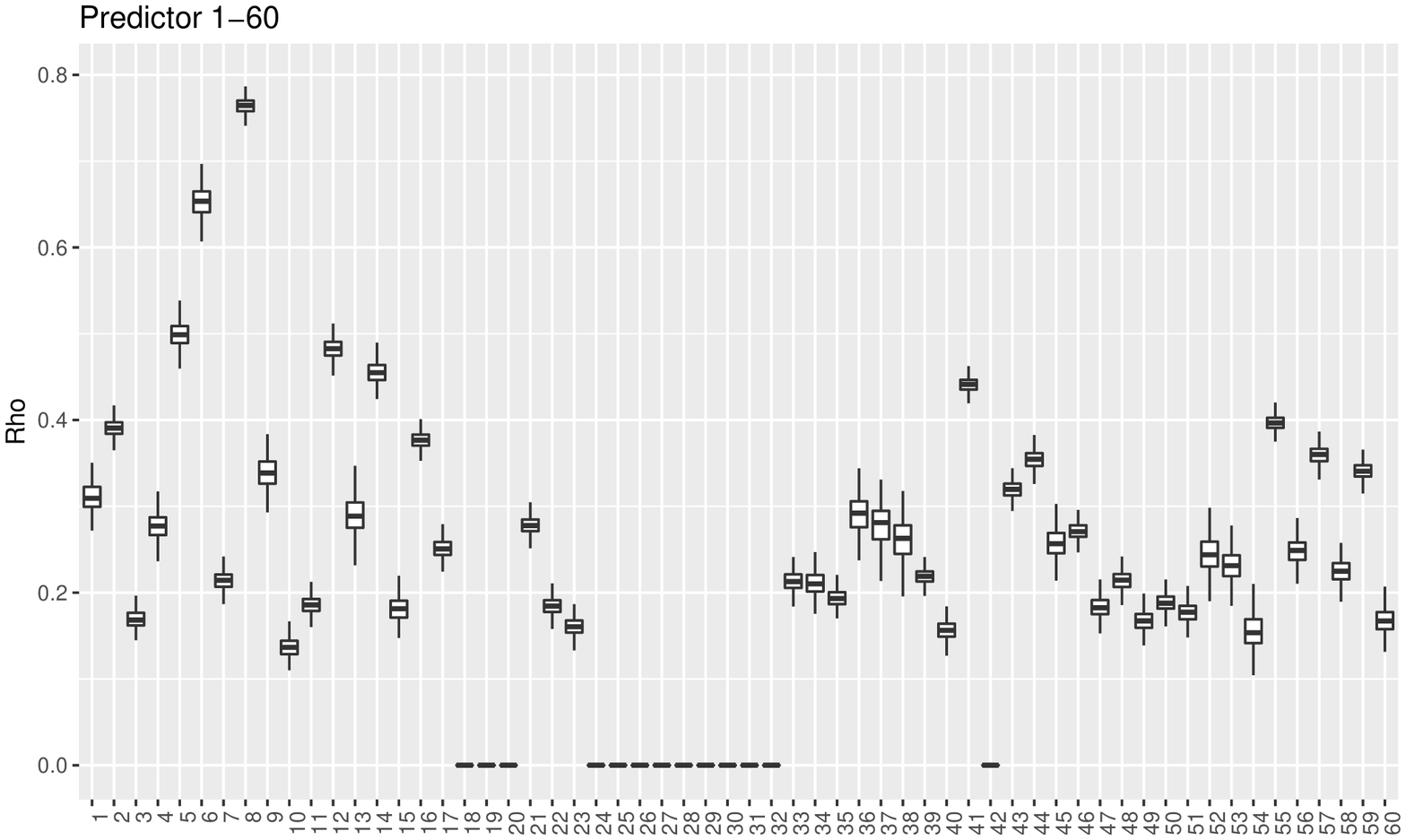}
\includegraphics[width=13cm,height=6.5cm]{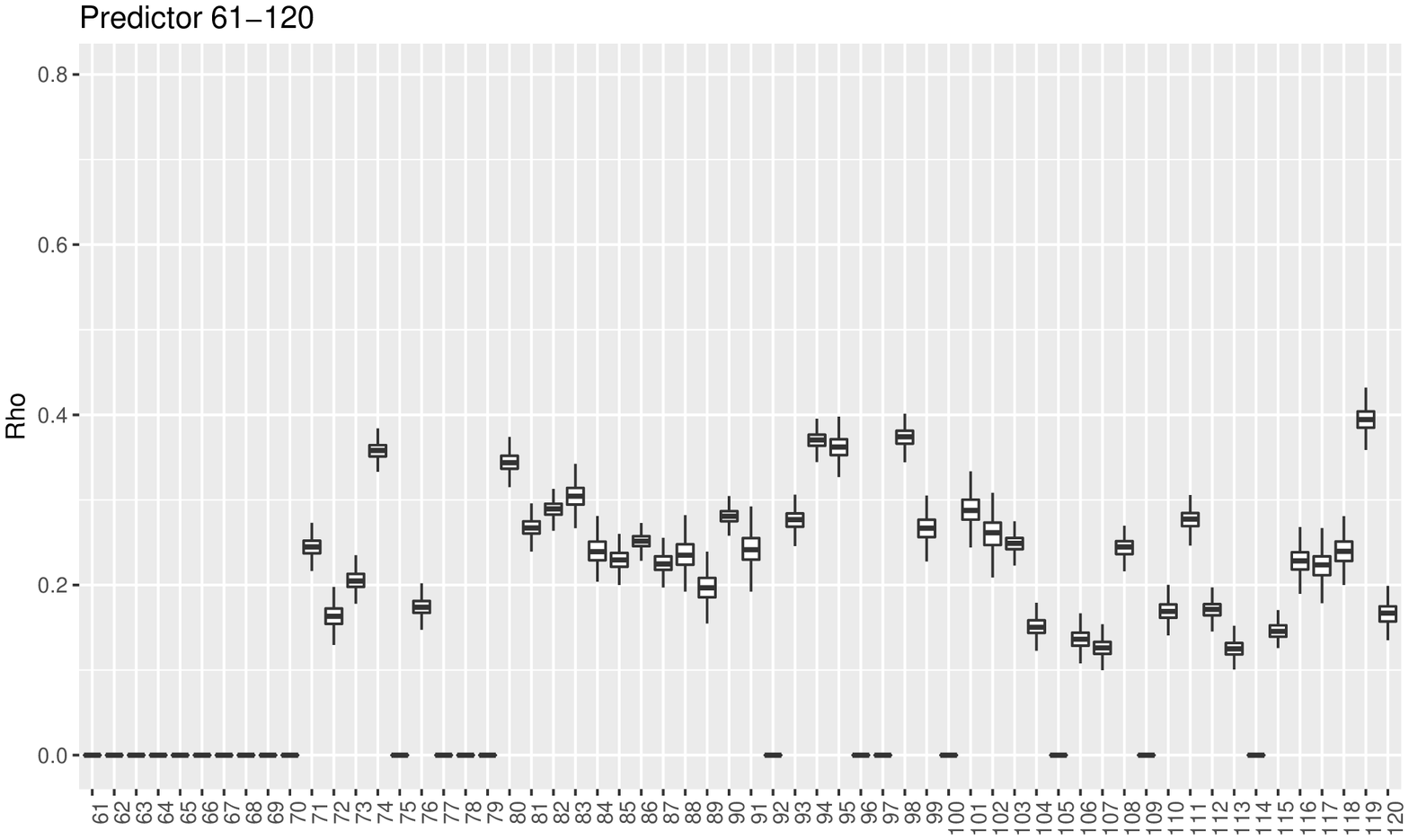}
\includegraphics[width=13cm,height=6.5cm]{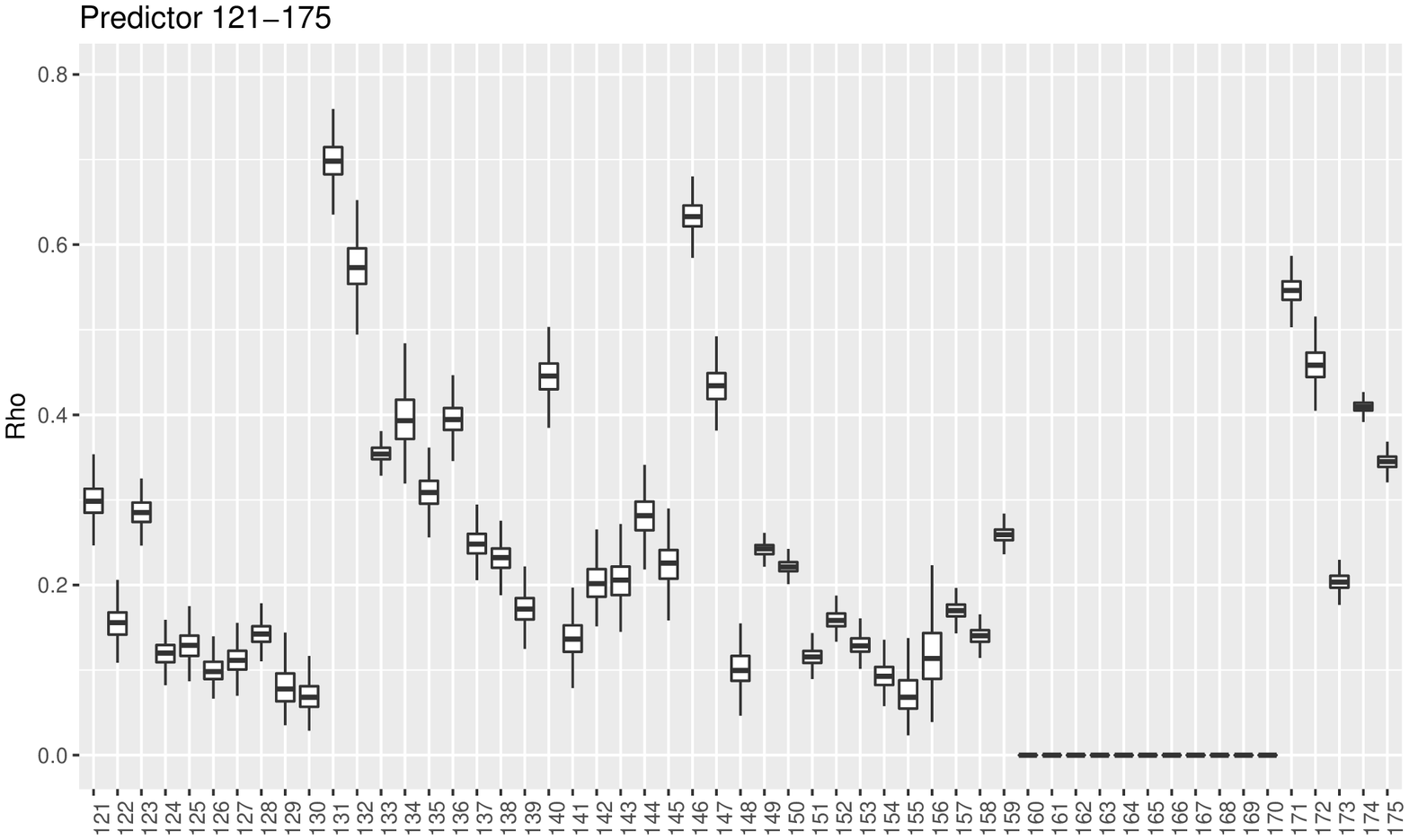}
\caption{Boxplots of $\delta$ for predictors 1-60 (above), 61-120 (middle), 121-175 (bottom). The horizontal and vertical axes represent the index and the estimated value of $\delta$ for each predictor. The predictors with more than 5\% missing rates are excluded from the analysis by setting zero.}
\label{fig:delta}
\end{center}
\end{figure}

\begin{figure}[htpb] 
\begin{center}
\hspace{-1cm}
\includegraphics[scale=0.85]{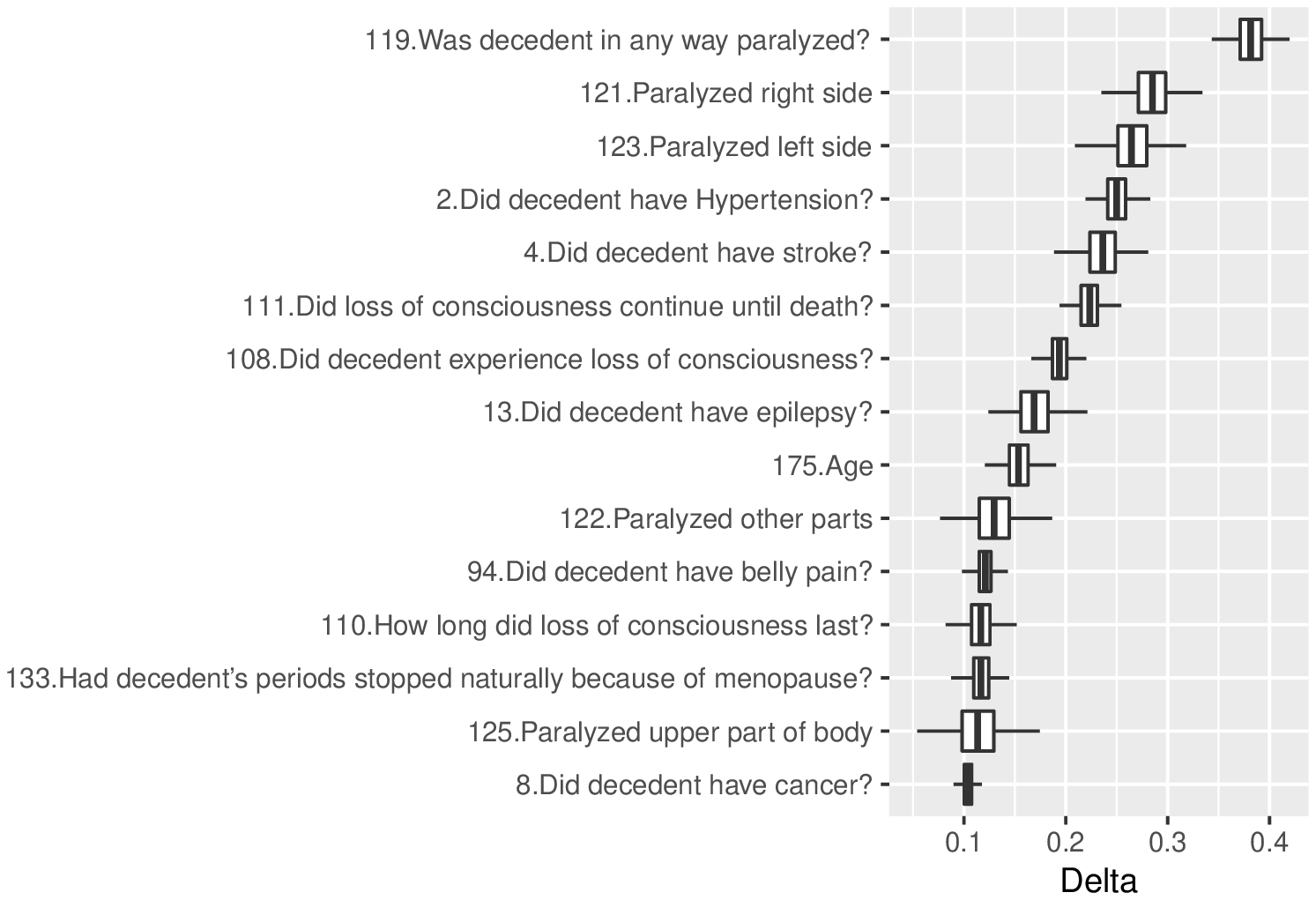}
\caption{Boxplot of $\delta$ for top 15 predictors in stroke in descending order of the posterior mean.}
\label{fig:measure-stroke}
\end{center}
\end{figure}

\begin{figure}[htpb] 
\begin{center}
\hspace{-1cm}
\includegraphics[scale=0.85]{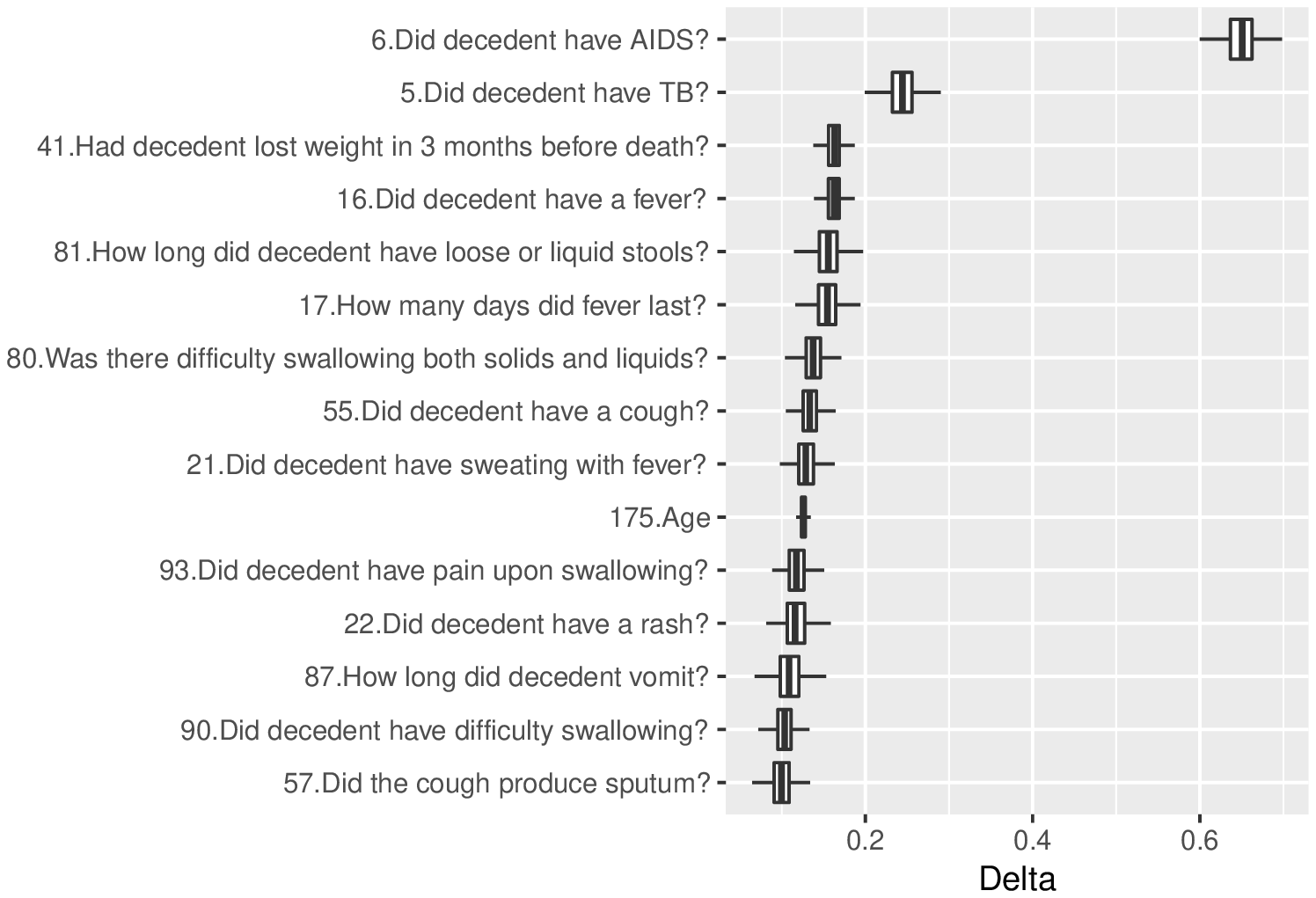}
\caption{Boxplot of $\delta$ for top 15 predictors in AIDS in descending order of the posterior mean.}
\label{fig:measure-AIDS}
\end{center}
\end{figure}

\bibliographystyle{chicago}
\bibliography{VA3,lancetVA,R01-VA}